\documentclass[a4paper,fleqn,usenatbib]{style/mnras}

\usepackage{newtxtext,newtxmath}

\usepackage[T1]{fontenc}
\usepackage{ae,aecompl}

\usepackage{graphicx}	
\usepackage{amsmath}	
\usepackage{amssymb}	

\usepackage{natbib}
\usepackage{booktabs}
\usepackage{longtable, booktabs}
\usepackage[titletoc,title]{appendix}
\usepackage{color}
\usepackage{xcolor}
\usepackage{lscape}

\usepackage{newtxtext,newtxmath}

\usepackage{comment} 

\newcommand{\cii}{[CII]}
\newcommand{\CII}{[CII]}

\def\msun{{\rm M}_{\odot}}
\def\zsun{{\rm Z}_{\odot}}

\def\msunyr{\msun/{\rm yr}}
\def\msunpc2{\msun/{\rm pc}^{2}}

\def\althaea{Alth{\ae}a}
\newcommand{\quotes}[1]{``#1''}
\newcommand\code[1]{\textsc{\MakeLowercase{#1}}}

\definecolor{mkcolor}{HTML}{006699} 
\definecolor{apcolor}{HTML}{b3003b}
\definecolor{afcolor}{HTML}{01bdff}

\graphicspath{{../figures/},{figures/}} 

\title[Galaxy growth:mergers or gravitational instability]{Early galaxy growth: mergers or gravitational instability?} 
\author[A. Zanella et al.]{
A. Zanella$^{1}$\thanks{E-mail: \href{anita.zanella@inaf.it}{anita.zanella@inaf.it}},
A. Pallottini$^2$, A. Ferrara$^2$, S. Gallerani$^2$, S. Carniani$^2$, M. Kohandel$^2$, 
\newauthor C. Behrens$^3$
\\
$^{1}$Istituto Nazionale di Astrofisica, Vicolo dell'Osservatorio 5, 35122 Padova (Italy)\\
$^2$ Scuola Normale Superiore, Piazza dei Cavalieri 7, I-56126 Pisa (Italy)\\
$^3$ Institut f\"ur Astrophysik, Georg-August Universit\"at G\"ottingen, Friedrich-Hundt-Platz 1, 37077, G\"ottingen (Germany)
}

\date{Accepted XXX. Received YYY; in original form ZZZ}

\pubyear{2020}

\begin{document}
\label{firstpage}
\pagerange{\pageref{firstpage}--\pageref{lastpage}}
\maketitle

\begin{abstract}
We investigate the spatially-resolved morphology of galaxies in the early Universe.
We consider a typical redshift $z = 6$ Lyman Break galaxy, ``{Alth\ae a}'' from the SERRA hydrodynamical simulations. We create mock rest-frame ultraviolet, optical, and far-infrared observations, and perform a two-dimensional morphological analysis to deblend the galaxy disk from substructures (merging satellites or star-forming regions).
We find that the \cii 158$\mu$m emitting region has an effective radius 1.5 -- 2.5 times larger than the optical one, consistent with recent observations. This \cii~halo in our simulated galaxy arises as the joint effect of stellar outflows and carbon photoionization by the galaxy UV field, rather than from the emission of unresolved nearby satellites.
At the typical angular resolution of current observations ($\gtrsim 0.15"$) only merging satellites can be detected; detection of star-forming regions requires resolutions of $\lesssim 0.05"$. The \cii-detected satellite has a 2.5 kpc projected distance from the galaxy disk, whereas the star-forming regions are embedded in the disk itself (distance $\lesssim 1$ kpc). This suggests that multi-component systems reported in the literature, which have separations $\gtrsim 2$ kpc, are merging satellites, rather than galactic substructures.
%
Finally, the star-forming regions found in our mock maps follow the local L$_\mathrm{\cii}$ -- SFR$_\mathrm{UV}$ relation of galaxy disks, although sampling the low-luminosity, low-SFR tail of the distribution. We show that future \textit{JWST} observations, bridging UV and \cii~datasets, will be exceptionally suited to characterize galaxy substructures thanks to their exquisite spatial resolution and sensitivity to both low-metallicity and dust-obscured regions that are bright at infrared wavelengths.
\end{abstract}




\section{Introduction}
\label{sec:intro}

Galaxies in the first billion years of the Universe lifetime undergo a rapid assembly phase and their properties quickly change over time. The period between redshift $z = 4 - 6$ represents a key transition phase from the primordial Universe, when neutral hydrogen was ionized by the first sources ($z > 6$), and the peak of cosmic star formation rate (SFR) density, when galaxies are mature ($ z \sim 2 - 3$). Studying galaxies at this early epoch is important to understand how they assemble their mass while building up the structures that are commonly observed in local sources (e.g. disk, bulge). In particular, analyzing their morphology provides key insights into their formation and structural evolution. 

In the local Universe most of the star-forming galaxies show a central bulge with old stellar populations embedded in a thick disk with spiral arms hosting giant molecular clouds (molecular gas mass $M_\mathrm{mol} \sim 10^5 - 10^7$ M$_\odot$) and star clusters (stellar mass M$_\star \sim 10^3 - 10^6$ M$_\odot$, \citealt{Conselice2014} and references therein). The bulk of star-forming galaxies at $z \sim 1 - 3$ instead shows irregular morphologies with thin disks dominated by massive star-forming regions (M$_\star \sim 10^7 - 10^9$ M$_\odot$, size $\lesssim 1$ kpc) with bright blue colors, typically called \quotes{clumps} (\citealt{Bournaud2016} and references therein). It is still debated whether clumps are transient features quickly disrupted by their own intense stellar feedback (\citealt{Genel2012}, \citealt{Hopkins2012}, \citealt{Moody2014}, \citealt{Tamburello2016}, \citealt{Oklopcic2017}) or if they survive for $\gtrsim 100$ Myr. In the latter case, they are expected to migrate inward, contribute to the formation of the bulge and the thickening of the disk, playing a key role in galaxy evolution (\citealt{Immeli2004a}, \citealt{Dekel2009}, \citealt{Ceverino2012}, \citealt{Inoue2016}, \citealt{Mandelker2015}, \citealt{Mandelker2017}).
Also the origin of clumps is still under investigation: it is unclear in fact whether they are remnants of merging satellites or if instead they formed in the gas-rich, turbulent disk of galaxies due to gravitational disk instability (\citealt{Bournaud2008}, \citealt{Genzel2008}, \citealt{Genzel2011}, \citealt{Puech2009}, \citealt{Wuyts2014}, \citealt{Guo2015}, \citealt{Zanella2015}, \citealt{Ribeiro2017}, \citealt{Fisher2017a}). Different avenues have been followed to tackle this problem, including the analysis of spatially-resolved velocity and Toomre parameter maps (\citealt{Forster-Schreiber2011b}, \citealt{Wisnioski2011}, \citealt{Genzel2014}, \citealt{Girard2018}); the characterisation of physical properties (e.g. size, mass, metallicity, stellar populations) of statistical samples of clumps \citep{Guo2018, Zanella2019}; the investigation of the redshift evolution of the number fraction of clumpy galaxies in the overall star-forming population \citep{Guo2015, Shibuya2016}. By comparing the fraction of clumpy galaxies at $z \sim 0 - 3$ with theoretical predictions, \cite{Guo2015} conclude that clumps in galaxies with M$_\star \gtrsim 10^{11}$ M$_\odot$ are likely merger remnants, whereas in lower mass galaxies they form through gravitational instability. \cite{Shibuya2016} extend this analysis to higher redshift, including a sample of Lyman break galaxies (LBGs) at $z \sim 4 - 8$ observed with the \textit{Hubble} Space Telescope (\textit{HST}). They find that the fraction of clumpy galaxies at $z \gtrsim 3$ decreases following the drop of the cosmic SFR density. They conclude that only the theoretical works predicting the \textit{in-situ} growth of clumps can simultaneously reproduce the fraction of clumpy galaxies observed at low- and high-redshift. Spatially-resolved observations of individual $z \sim 4 - 6$ galaxies however are needed to determine the physical properties of clumps in the early Universe, confirm their \textit{in-situ} origin, and understand what is their role in the evolution of the host galaxy.
Furthermore, to build a comprehensive picture of galaxy formation, it is important to analyze multiwavelength datasets. Rest-frame ultraviolet (UV) observations tracing unobscured star formation and stellar winds (\citealt{Heckman1997}, \citealt{Maraston2009}, \citealt{Steidel2010}, \citealt{Faisst2016a}) are complementary to far-infrared (FIR) continuum and emission line (e.g. \cii) data tracing the obscured star formation, gas and dust content (\citealt{deLooze2014}, \citealt{Pavesi2019}). In recent years, the exquisite resolution and sensitivity of the Atacama Large Millimeter / submillimeter Array (ALMA) made possible the comparison of the FIR morphology of $z \sim 4 - 6$ galaxies with the UV one shown by \textit{HST}. These studies revealed a large fraction of multi-component systems with complex morphology where, in some cases, the UV and FIR emission are even spatially offsetted (\citealt{Carniani2018}, \citealt{Lefevre2019}, \citealt{Jones2020} and references therein). The origin of these substructures is unclear: they could be mergers or galaxies with massive clumps. A detailed multiwavelength study of these systems is important to understand their nature, the fraction of dust-obscured satellites and clumps (that would be missed in UV-based surveys), and in turn understand the contribution of gravitational instability and mergers to galaxy mass assembly and evolution.

As observations progressed, several models of galaxies in the epoch of reionization were developed. In particular cosmological simulations have been used to zoom-in on the structure of high-redshift galaxies and investigate their contribution to the reionization (\citealt{Katz2017}, \citealt{Trebitsch2017}, \citealt{Rosdahl2018}, \citealt{Hopkins2018a}), as well as the role of stellar feedback in the formation and evolution of these primordial sources (\citealt{Agertz2015}, \citealt{Pallottini2017a}). Simulations have been key also to investigate the chemical enrichment processes in galaxies at high redshift (\citealt{Maio2016}, \citealt{Smith2017}, \citealt{Pallottini2017b}, \citealt{Lupi2018}, \citealt{Capelo2018}), and the dust content of such sources \citep{Behrens2018,liang:2019}.

With this work we aim at bridging simulations and observations to investigate how primeval galaxies form and assemble their mass. Starting from the zoomed-in cosmological simulations of Lyman break galaxies developed by \citealt{Pallottini2017b}, we create mock rest-frame ultraviolet, optical, and far-infrared observations with the goal of analyzing the structure of these galaxies and relate it to the morphology of $z \sim 6$ observed sources. In particular we compare two stages: in the first the galaxy appears as an undisturbed clumpy disk, in the second it is undergoing a merger. We investigate how the galaxy morphology differs in these two cases and what components are detected when considering different tracers (e.g. rest-frame UV and FIR emission).
This paper is organized as follows: in Section \ref{sec:simulations} we summarize the main characteristics of the zoom-in cosmological simulations adopted in this work; in Section \ref{sec:mock_obs} we describe how we created mock rest-frame ultraviolet, optical, and far-infrared observations; in Section \ref{sec:analysis} we discuss how we analyzed the mock two-dimensional maps and we measured the structural properties of galaxies; in Section \ref{sec:results} we report the results of the analysis; in Section \ref{sec:discussion} we interpret our results and compare them with those reported in the literature; finally in Section \ref{sec:conclusions} we summarize and conclude.
Throughout the paper we use a flat $\Lambda$CDM cosmology with $\Omega_\mathrm{m} = 0.3$, $\Omega_\mathrm{\Lambda} = 0.7$, and H$_0 = 70$ km s$^{-1}$ Mpc$^{-1}$. We assume a \cite{Kroupa2001} initial mass function (IMF) and, when necessary, we accordingly converted literature results obtained with different IMFs.

\section{Galaxy simulations}
\label{sec:simulations}

The adopted hydrodynamical simulations are fully described in \citet{Pallottini2017b}. The simulation is based on a modified version of the adaptive mesh refinement code \code{RAMSES}\footnote{\url{https://bitbucket.org/rteyssie/ramses/}} \citep{Teyssier2002}, in order to evolve a comoving cosmological volume of $(20 {\rm Mpc}/h)^3$, that is generated with \code{music}\footnote{\url{https://bitbucket.org/ohahn/music/}} \citep{hahn:2011mnras}. The simulation zooms-in the Lagrangian region of of a dark matter halo of mass $\simeq 3.5\times 10^{11} \msun$ at $z \simeq 6$, that hosts the galaxy \quotes{\althaea}. The gas mass resolution in the zoomed region is $10^4 \msun$ and at $z=6$ it is resolved to spatial scales of $\simeq$ 30 pc\footnote{The simulation adopts a fixed resolution in comoving coordinates, thus the physical resolution degrades as the simulation evolve in time, reaching the worst resolution (30 pc) at $z=6$.}, by adopting a quasi-Lagrangian mass-refinement criterion.
In the simulation stars are formed from molecular hydrogen according to a Schmidt-Kennicutt relation \citep{Schmidt1959,Kennicutt1998}. The abundance of the molecular hydrogen is computed using time dependent chemical network implemented using the \code{KROME}\footnote{\url{https://bitbucket.org/tgrassi/krome}} package \citep{Grassi2014,Bovino2016}. Stellar feedback includes supernova explosions, radiation pressure, and winds from massive stars \citep{Pallottini2017b}; the model also accounts for the blastwave propagation inside molecular clouds, and the thermal and turbulent energy content of the gas is modelled similarly to \citet{Agertz2015}. Stellar energy inputs and chemical yields are calculated via \code{STARBURST99} \citep{Leitherer1999} assuming a \citet{Kroupa2001} IMF for the star clusters. In this simulation a spatially uniform interstellar radiation field is considered, and its intensity scales with the star formation rate of the galaxy \citep[see for details][]{pallottini:2019}.

\althaea~appears as a typical $z \geq 6$ LBG \citep{Behrens2018,behrens:2019}, following the SFR -- $M_\star$ relation observed at high-$z$ \citep{jiang:2016}. At the earliest epochs, it is constituted by a small disk surrounded by several substructures (size $< 100$ pc) -- typically coinciding with molecular cloud complexes \citep{leung:2019} -- and it is fed with gas through filaments \citep{Kohandel2019}. As time passes, the disk grows in size and mass thanks to \textit{in-situ} star-formation and mergers with satellites which are disrupted and embedded in the disk \citep{gelli:2020}.

For this work we focus on two specific evolutionary stages with a morphological distinct structure \citep[see also][]{Kohandel2019,kohandel:2020}: (a) a \textit{clumpy disk}, found at $z = 7.2$, with a total stellar (molecular gas) mass $M_\star = 7.1 \times 10^9 M_\odot$ ($M_{\rm H2}= 2.0 \times 10^7 M_\odot$), a star formation rate\footnote{In the present work, the star formation rate is computed accounting for stars with age $t_\star < 30 \, \rm Myr$.} SFR $= 49.2 M_\odot$ yr$^{-1}$, and metallicity $Z= 0.8 Z_\odot$; (b) a \textit{merger}, found at $z = 6.47$, with a total stellar (molecular gas) mass $M_\star = 9.9 \times 10^9 M_\odot$ ($M_{\rm H2}= 2.4 \times 10^7 M_\odot$), a star formation rate SFR $= 48.1 M_\odot$ yr$^{-1}$, and metallicity $Z= 0.7 Z_\odot$ (Table \ref{tab_ph_properties}). For a fair observational comparison, we redshifted both of them to $z=6$. In the rest of the paper we will consider these two stages as independent sources.

\begin{table}
    \caption{Physical properties of {Alth\ae a} in the two stages considered for our analysis: the clumpy disk and the merger with a nearby satellite.}
    \label{tab_ph_properties}
\resizebox{\columnwidth}{!}{%
\begin{tabular}{ccccccc}
\toprule
\midrule
         & $M_\star$              & ${\rm SFR}$                      & $M_g$                 & $M_{\rm H2}$ & $Z$   &   $A_V$   \\
~            & $10^9 {\rm M}_{\odot}$ & ${\rm M}_{\odot}\,{\rm yr}^{-1}$ & $10^9{\rm M}_{\odot}$ & $10^7{\rm M}_{\odot}$ & $Z_\odot$ &  \\
\midrule
Clumpy disk      & $7.1$                 & $49.2$                          & $1.7$                & $2.0$ & 0.8    & 1.5           \\
Merging galaxy      & $9.9$                 & $48.1$                          & $1.8$                & $2.4$   & 0.7  & 0.9            \\
Satellite & $1.3$                  & $10.8$                          & $3.4$                & $0.4$   & 1.1     & 1.5        \\
\bottomrule
\end{tabular}
}
\begin{minipage}{0.5\textwidth}
\textbf{Columns:} (1) Galaxy. (2) Stellar mass. (3) Star formation rate. (4) Total gas mass. (5) Molecular gas mass. (6) Metallicity. (7) Extinction.
\end{minipage}
\end{table}

\section{Multi-wavelength mock maps}
\label{sec:mock_obs}

To compare the morphology and structural parameters of our simulated galaxies with actual $z \sim 5 - 7$ observations, the first step is to create synthetic continuum and emission line maps. In particular we aim at reproducing typical \textit{HST} optical images (bands $z'$, $Y$, $J$, and $H$), ALMA sub-millimeter continuum (Band 6) and emission line (\cii) two-dimensional (2D) maps, as well as realistic near- and mid-infrared \textit{JWST} observations.

First, starting from the simulated galaxy, we generated mock continuum and emission line images by using \code{skirt} \citep{camps:2015} and \code{CLOUDY} \citep{ferland:2017} respectively (Section \ref{subsec:emissionmodel}). Then we added observational artifacts to mimic typical UV and FIR high-redshift observations (Sections \ref{subsec:optical_submm_mock}).

\subsection{Continuum and emission lines modelling}\label{subsec:emissionmodel}

Continuum emission is generated by using \code{skirt}\footnote{version 8.0, \url{http://www.skirt.ugent.be}} \citep{baes:2015,camps:2015}, a Monte Carlo based code that computes the radiative transfer process in dusty media. The setup adopted here is similar to \citet{Behrens2018}, and we summarize it as follows.
The spatial distribution of the light sources is taken from the position of the stellar clusters in \althaea; for each cluster, we use its metallicity and age to compute the stellar spectral energy distribution (SED), by adopting the \citet{Bruzual2003} models, and the same \citet{Kroupa2001} IMF used in the simulation.

In \citet{Pallottini2017b}, the metal content of the gas is evolved accounting for supernovae and processed ejecta from stellar winds, starting from a metallicity $Z=10^{-3}\zsun$ floor, as expected from a pre-enrichment scenario \citep{tornatore:2007,pallottini:2014,maiolino:2019}. Dust is not directly traced and we adopt dust-to-metal ratio $f_\mathrm{d}=0.08$. such value is found in \citet{Behrens2018} in order to have an observed UV and FIR SED comparable to high-redshift observations \citep{laporte:2017}. Note that in the Milky Way $f_\mathrm{d} = 0.3$ and typically $f_\mathrm{d} \simeq 0.2$ in local galaxies \citep{delooze:2020}, while at high-redshift the value is much more uncertain \citep{wiseman:2017}. Dust composition and grain size distribution is set to mimic the Milky-Way \citep{weingartner:2001}, and we assume a dust emissivity $\beta_d = 2$.

We use \code{CLOUDY}\footnote{C17.01 \url{https://www.nublado.org/}} \citep{ferland:2017} to compute the line emission for \CII~and CO roto-vibrational transitions.
Similarly to \citep{pallottini:2019}, we use grids of \code{cloudy} models for density, metallicity, radiation field intensity, as a function of the column density. We account for the turbulent and clumpy structure of the interstellar medium (ISM), by parameterizing the underlying distribution as a function of the gas Mach number \citep{vallini:2017,vallini:2018}.
With respect to \cite{pallottini:2019}, here the radiation field is assumed to be uniform and non-ionizing, thus photoevaporation effects are not fully included: such effect can both modify the emission lines strength \citep{vallini:2017} and the H$_2$ -- and thus star -- formation \citep{decataldo:2019}. While in general line emission is sensitive to adopted models \citep{olsen:2018}, the resulting \cii~flux is more robust to changes of assumptions \citep{lupi:2020}. Note that the considered FIR lines are optically thin given the column densities found in the ISM of \althaea, thus no further dust attentuation is needed. For a fair comparison, we set the spatial resolution of the continuum and lines at 25 pc for both the clumpy disk and merger case, and we redshifted both of them to $z=6$. In both snapshots the galactic disk is seen face-on.

\begin{center}
\begin{table}
    \caption{Parameters of the mock observations}
\resizebox{\columnwidth}{!}{%
\begin{tabular}{c c c c c}
\toprule
\midrule
Telescope     & Band                         & $\lambda_{\mathrm{c}}$   & Angular resolution & Depth \\
              &                              & ($\mu$m)                  & (arcsec$^2$)       &       \\
(1)           & (2)                          & (3)                    & (4)                &   (5)  \\
\midrule
\textit{HST}  & ACS/F850LP (\textit{z'})     & 0.9   & $0.13 \times 0.13$ & 29.0 \\
              &                              &       & $0.04 \times 0.04$ & 29.0 \\
\textit{HST}  & WFC3/F105W (\textit{Y})      & 1.1   & $0.13 \times 0.13$ & 29.0 \\
              &                              &       & $0.04 \times 0.04$ & 29.0 \\
\textit{HST}  & WFC3/F125W (\textit{J})      & 1.2   & $0.13 \times 0.13$ & 29.0 \\
              &                              &       & $0.04 \times 0.04$ & 29.0 \\
\textit{HST}  & WFC3/F160W (\textit{H})      & 1.5   & $0.15 \times 0.15$ & 29.0 \\
              &                              &       & $0.05 \times 0.05$ & 29.0 \\
\textit{JWST} & NIRCam/F444W (\textit{near-IR}) & 4.4 & $0.15 \times 0.15$ & 29.0 \\
              &                                 &     & $0.05 \times 0.05$ & 29.0 \\
\textit{JWST} & MIRI/F770W (\textit{mid-IR})    & 7.7 & $0.24 \times 0.24$ & 29.0 \\
              &                                 &     & $0.08 \times 0.08$ & 29.0 \\
ALMA          & Band 6 (continuum)              & 1100.0 & $0.18 \times 0.12$ & 3.4  \\
              &                                 &      & $0.05 \times 0.04$ & 4.7  \\
ALMA          & Band 6 ([CII])                  & 1100.0 & $0.18 \times 0.12$ & 19.0 \\
              &                                 &      & $0.05 \times 0.04$ & 18.0 \\
\bottomrule
\end{tabular}
}
\begin{minipage}{0.5\textwidth}
\textbf{Columns:} (1) Telescope. (2) Camera and filter. (3) Central wavelength of the filter. (4) Angular resolution achieved in the mock observations for the low- (top) and high-resolution (bottom) case. See Section \ref{sec:mock_obs} for details. (5) Depth of the observations. We report the limiting magnitude (in AB mag) defined as 5$\sigma$ sky noise in a 0.25"-radius aperture for the \textit{HST} and \textit{JWST} observations. We report the rms noise level (in $\mu$Jy beam$^{-1}$) for the ALMA observations.
\end{minipage}
    \label{tab:observations}
\end{table}
\end{center}

\begin{figure*}
\centering
\includegraphics[width=1.15\textwidth, angle=90]{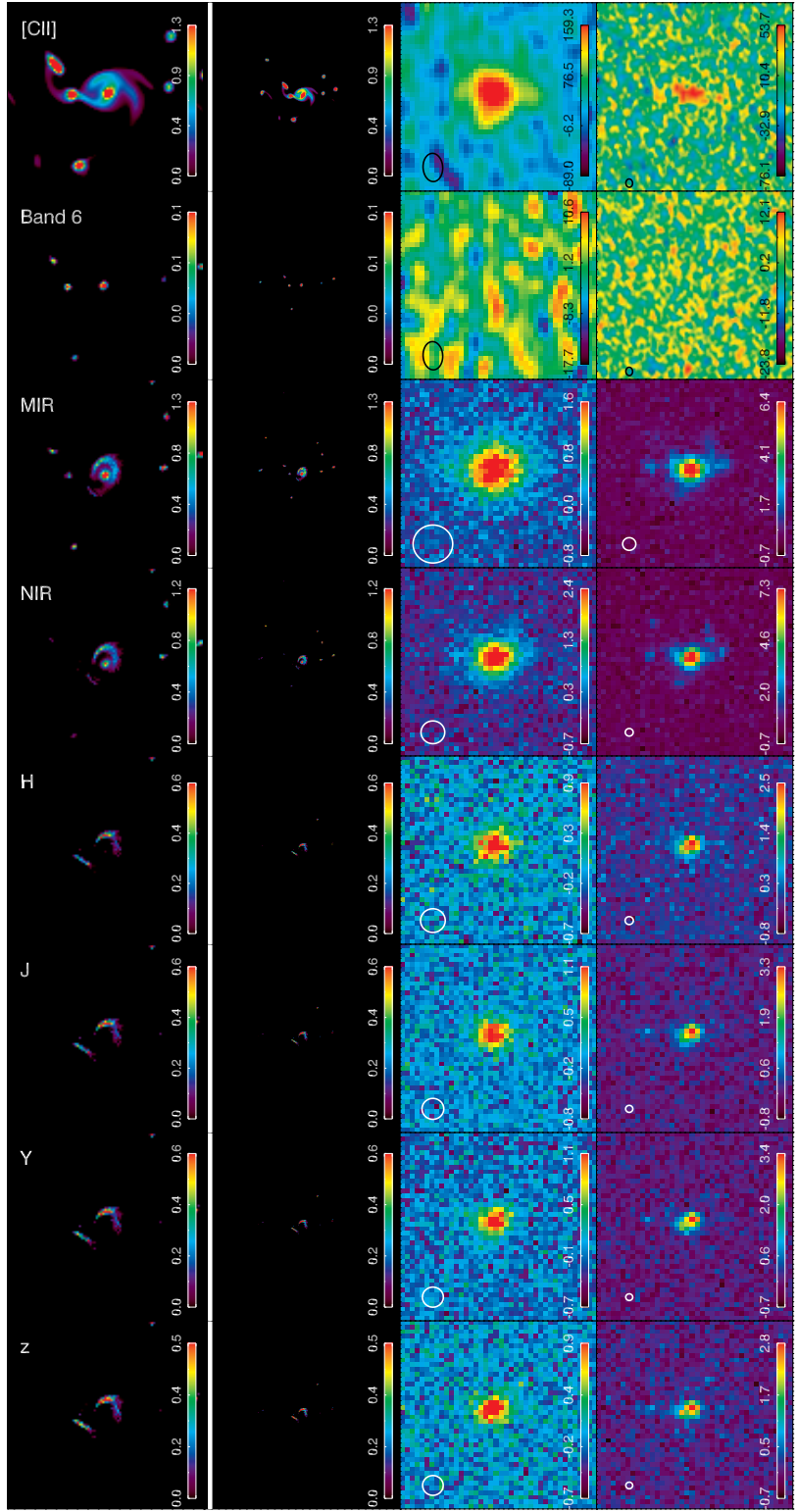}
\caption{Maps of {Alth\ae a} in the clumpy disk stage. \textbf{From top to bottom:} different observing bands are shown. \textit{HST}/ACS F850LP (\textit{z'}), \textit{HST}/WFC3 F105W (\textit{Y}, F125W (\textit{J}), F160W (\textit{H}); \textit{JWST}/NIRCam F444W (\textit{NIR}), \textit{JWST}/MIRI F770W (\textit{MIR}); ALMA Band 6 continuum and the [CII] pseudo-narrow band emission line map. \textbf{From left to right:} maps with different spatial resolutions are shown, in particular the nominal resolution from the simulation (first two panels), the typical resolution of current observations ($\sim$ 0.15'' - 0.2'', third panel), and the higher resolution currently achievable with ALMA and/or in lensed systems ($\sim$ 0.05'', considering a magnification factor $\mu \sim 10$, fourth panel). The stamps in the leftmost column have a size of 0.2'' $\times$ 0.2'' ($\sim 1.1 \times 1.1$ kpc at $z \sim 6$), whereas the other stamps have a size of 0.6'' $\times$ 0.6'' ($\sim 3.4 \times 3.4$ kpc at $z \sim 6$).  The color bars report the flux values in units of $10^{-3} \mu$Jy for the \textit{HST} and \textit{JWST} maps, and $\mu$Jy beam$^{-1}$ for the ALMA maps. The white and black circles in the bottom left corner of the maps indicate the spatial resolution of the observations (the full width at half maximum of the point spread function for the \textit{HST} and \textit{JWST} maps, the beam for the ALMA bands).}
\label{fig:maps16}
\end{figure*}

\begin{figure*}
\centering
\includegraphics[width=1.25\textwidth, angle=90]{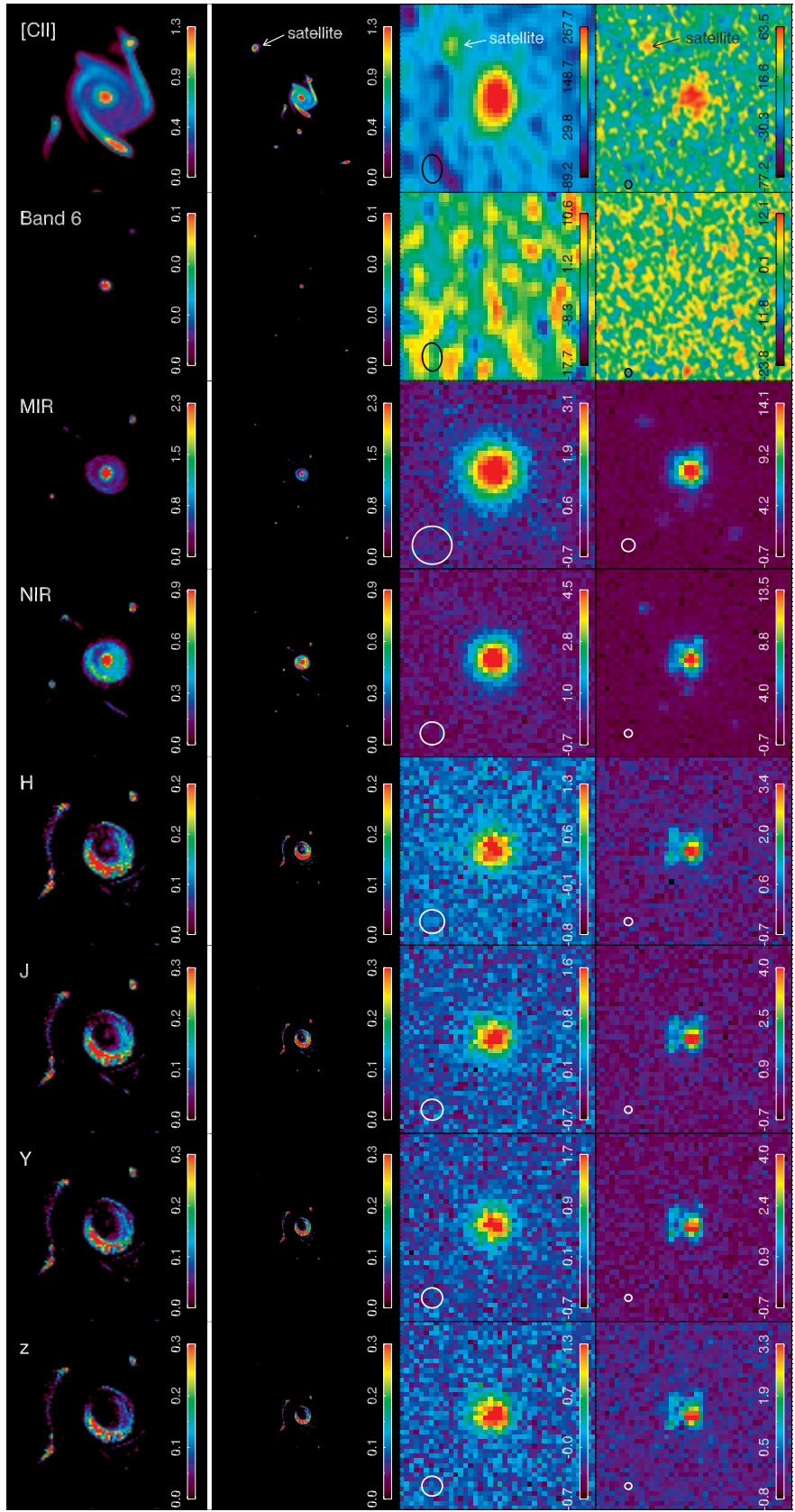}
\caption{Maps of {Alth\ae a} in the merger stage. Colors and symbols are as in Figure \ref{fig:maps16}.}
\label{fig:maps23}
\end{figure*}

\begin{center}
\begin{table*}
    \caption{Structural parameters.}
\resizebox{\textwidth}{!}{%
\begin{tabular}{c c c c c c c c c c}
\toprule
\midrule
Galaxy    & Resolution & Component  & F$_{\textit{z'}}$ & F$_{\textit{Y}}$ & F$_{\textit{J}}$ & F$_{\textit{H}}$ & F$_{\mathrm{near-IR}}$ & F$_{\mathrm{mid-IR}}$ & F$_{\mathrm{[CII]}}$ \\
          &        &     &    ($\mu$Jy) &  ($\mu$Jy) & ($\mu$Jy)    & ($\mu$Jy) & ($\mu$Jy) & ($\mu$Jy) & ($\mu$Jy)  \\ 
(1)       & (2)         & (3)       & (4)       & (5)  & (6) & (7) & (8) & (9) & (10) \\
\midrule
Clumpy & Low-resolution & Disk & $0.051 \pm 0.010$ & $0.057 \pm 0.010$ & $0.063 \pm 0.010$ & $0.059 \pm 0.010$ & $0.210 \pm 0.010$ & $0.261 \pm 0.010$ & $614.3 \pm 131.1$ \\ 
         & High-resolution & Disk & $0.052 \pm 0.006$ & $0.054 \pm 0.006$ & $0.057 \pm 0.005$ & $0.055 \pm 0.005$ & $0.189 \pm 0.011$ & $0.244 \pm 0.003$ & $< 806.0$ \\
         &       & Substructure 1 & $< 0.003$ & $< 0.003$ & $< 0.003$ & $< 0.003$ & $0.005 \pm 0.001$ & $< 0.009$ & $< 29.0$ \\
         &       & Substructure 2 & $< 0.035$ & $< 0.041$ & $< 0.041$ & $< 0.039$ & $< 0.026$ & $0.189 \pm 0.011$ & $< 251.2$ \\
Merger & Low-resolution & Disk & $0.095 \pm 0.021$ & $0.115 \pm 0.018$ & $0.129 \pm 0.019$ & $0.109 \pm 0.022$ & $0.384 \pm 0.030$ & $0.506 \pm 0.095$ & $787.8 \pm 144.8$ \\
          &         & Substructure & $< 0.005$ & $< 0.003$ & $< 0.005$ & $< 0.004$ & $0.009 \pm 0.003$ & $< 0.027$ & $106.7 \pm 17.8$ \\
          & High-resolution & Disk & $0.078 \pm 0.006$ & $0.091 \pm 0.006$ & $0.102 \pm 0.006$ & $0.095 \pm 0.005$ & $0.360 \pm 0.008$ & $0.466 \pm 0.003$ & $567.0 \pm 135.0$ \\ 
          &       & Substructure 1 & $0.005 \pm 0.001$ & $0.006 \pm 0.001$ & $0.007 \pm 0.001$ & $0.007 \pm 0.001$ & $0.007 \pm 0.001$ & $0.010 \pm 0.003$ & $< 28.6$ \\
          &       & Substructure 2 & $0.006 \pm 0.001$ & $0.007 \pm 0.001$ & $0.007 \pm 0.001$ & $0.007 \pm 0.001$ & $0.007 \pm 0.001$ & $0.010 \pm 0.003$ & $< 28.8$ \\
          &       & Substructure 3 & $< 0.004$ & $< 0.004$ & $< 0.004$ & $< 0.003$ & $0.008 \pm 0.001$ & $0.012 \pm 0.003$ & $32.0 \pm 9.6$ \\
          &       & Substructure 4 & $< 0.004$ & $< 0.004$ & $< 0.004$ & $< 0.003$ & $0.008 \pm 0.001$ &  $0.013 \pm 0.002$ & $49.4 \pm 13.9$ \\
\midrule
\midrule
Galaxy    & Resolution & Component  & R$_{\mathrm{e,}\textit{z'}}$ & R$_{\mathrm{e,}\textit{Y}}$ & R$_{\mathrm{e,}\textit{J}}$ & R$_{\mathrm{e,}\textit{H}}$ & R$_{\mathrm{e,}\mathrm{near-IR}}$ & R$_{\mathrm{e,}\mathrm{mid-IR}}$ & R$_{\mathrm{e,}\mathrm{[CII]}}$ \\
          &       &      &    (pc) & (pc) & (pc) & (pc)    & (pc) & (pc) & (pc) \\ 
\midrule
Clumpy disk & Low-resolution & Disk & $246 \pm 37$ & $264 \pm 40$ & $278 \pm 42$ & $274 \pm 41$ & $442 \pm 88$ & $456 \pm 91$ & $733 \pm 147$ \\ 
         & High-resolution & Disk & $303 \pm 45$ & $300 \pm 45$ & $295 \pm 44$ & $293 \pm 44$ & $350 \pm 52$ & $346 \pm 52$ & $751 \pm 150$ \\
Merger & Low-resolution & Disk & $398 \pm 60$ & $416 \pm 62$ & $423 \pm 63$ & $363 \pm 54$ & $399 \pm 80$ & $398 \pm 80$ & $629 \pm 126$ \\
          & High-resolution & Disk & $295 \pm 44$ & $300 \pm 45$ & $308 \pm 46$ & $326 \pm 49$ & $348 \pm 70$ & $284 \pm 57$ & $536 \pm 107$ \\ 

\bottomrule
\end{tabular}
}
\begin{minipage}{0.96\textwidth}
\textbf{Columns:} (1) Galaxy: clumpy disk or merger. (2) Resolution: low- ($\sim 0.15"$) or high-resolution ($\sim 0.05"$). (3) Component: galaxy disk or substructure. (4) -- (10) Flux or effective radius of the disk measured in the given band.
\end{minipage}    
\label{tab:fluxes_sizes}
\end{table*}
\end{center}

\subsection{Mimicking observational artefacts}
\label{subsec:optical_submm_mock}

To properly reproduce actual observations we also need to mimic the image broadening due to the limited spatial resolution (i.e. diffraction limit), the pixelization of the detectors, and the presence of noise that limits the depth of the data.

We reproduced the case of spatially-resolved observations of galaxies where a typical image quality of $0.5 - 1$ kpc, corresponding to $0.1" - 0.2"$ at $z \sim 6$, is reached (depending on the observing band, e.g. \citealt{Grogin2011}, \citealt{Shibuya2015}). We also considered the ideal case of a galaxy observed with a resolution of $\sim$ 0.25 kpc, corresponding to $\sim 0.05"$ at this redshift. Currently this resolution is beyond the diffraction limit of \textit{HST} and is only achievable in moderately lensed sources (e.g. with magnification $\mu \gtrsim 10$, \citealt{Knudsen2016}, \citealt{Bradac2017}), or in the submillimeter with ALMA. However in the near future, the bluest \textit{JWST}/NIRCam filters at wavelength $\lambda \lesssim 1.5 \mu$m (corresponding to the \textit{HST} ones considered in this paper) will allow the community to achieve such a high spatial resolution also in the optical.
Throughout the paper we will refer to the maps with $\sim 0.15"$ resolution as the \quotes{low-resolution} case and to the ones with $\sim 0.05"$ resolution as the \quotes{high-resolution} case. 

We adopted the two following procedures to create mock maps in the optical and infrared (\textit{HST}, \textit{JWST}), and at sub-millimeter wavelengths (ALMA).
To obtain the final \textit{HST} and \textit{JWST} images we smoothed the original-resolution maps with a Gaussian kernel (Table \ref{tab:observations}).
To account for the pixelization of the detector we resampled the smoothed maps to a pixel scale of 0.03'' pixel$^{-1}$ that can be achieved in \textit{HST} imaging when dithering (e.g. \citealt{Zanella2019}). Adopting a larger pixel scale as in some studies (e.g. 0.06'' pixel$^{-1}$, \citealt{Brammer2012}), does not affect our results.
Finally, we have added random noise to reproduce the sensitivity of data taken in the commonly observed cosmological fields (e.g. CANDELS survey, \citealt{Grogin2011}). The 5$\sigma$ limiting magnitude of our mock optical and infrared images is $\sim 29$ AB mag considering a point-source and an aperture with $\sim 0.25"$ radius. The original-resolution images as well as the mock maps obtained for our clumpy disk and merger are shown in Figures \ref{fig:maps16} and \ref{fig:maps23}.

To reproduce the limited angular resolution, pixelization and noise of sub-millimeter images instead we used \code{CASA}\footnote{v5 \url{https://casa.nrao.edu/}}, the observing simulator of ALMA \citep{casa:2007ASPC}. We gave as input the original-resolution continuum and emission line models and generated $uv$ data with the \texttt{simobserve} task. Since in this work we do not focus on the kinematical properties of galaxies, but rather on their morphology, we did not produce hyperspectral cubes (including spatial and velocity information). We directly fed \code{CASA} with the 2D \cii~models integrated over a line width of $100$ km s$^{-1}$ \citep[cfr with][]{kohandel:2020}.
We then imaged the simulated observations with the \texttt{simanalyze} task. We adopted different configurations, in order to achieve $\sim 0.15"$ and $\sim 0.05"$ angular resolutions (Table \ref{tab:observations}). We set the observing time to $10$ hours, as this is a typical integration time for high-redshift observations (e.g. \citealt{Jones2017}, \citealt{laporte:2017}, \citealt{Carniani2018}). Imaging was performed using a Briggs weighting scheme (ROBUST = -0.5) which gives a good trade-off between resolution and sensitivity. 


\section{Morphological study}
\label{sec:analysis}

The original-resolution maps of our clumpy galaxy and merger do not simply appear as smooth disks, but they rather show several substructures (Figures \ref{fig:maps16}, \ref{fig:maps23}). In this work we aim at understanding whether these structures are still detected in realistic mock images that include the effects of limited spatial resolution and sensitivity. 

We also investigate how the structural parameters of the galaxy disk (e.g. effective radius, S\'ersic index) depend on the resolution and observing band used to measure them. In this Section we discuss the method that we used to deblend the substructures from the underlying disks, how we measured their continuum and emission line fluxes (with associated uncertainties) as well as the properties of the disks. We call ``substructures''  all the significant detections that depart from a smooth stellar disk (e.g. merging satellites, star-forming clumps), similarly to \cite{Guo2018} and \cite{Zanella2019}.


\subsection{Flux measurements}
\label{subsec:flux_measurements}

The procedure that we use to detect the substructure and disentangle it from the underlying galaxy disk is analogous to that described in \cite{Zanella2019}. In brief, we modelled with \code{GALFIT} \citep{Peng2010} the 2D light profile of the optical, infrared, and sub-millimeter maps independently. We adopted a single S\'ersic profile and we subtracted the best-fit model from each map. With this first step we could understand whether the galaxy disk could be considered smooth or if additional structures would appear in the residuals. By using \code{SExtractor} \citep{Bertin1996} independently on each residual map, we identified additional substructures and matched their coordinates. We considered that two detections were matched if their offset in different maps were smaller than the FWHM of the PSF ($\sim 0.15"$ in the low-resolution case, $0.05"$ in the high-resolution case). We created a final catalog with the coordinates of all the identified substructures.

We estimated the flux of the substructures and diffuse disk in the different continuum and \cii~maps as follows. We fitted again the 2D light profile of our galaxies considering simultaneously a S\'ersic profile to model the disk component and additional PSF or S\'ersic profiles at the location of the substructures detected with \code{SExtractor}. We considered S\'ersic instead of PSF profiles only if the substructure is resolved (i.e. its effective radius is larger than the FWHM of the PSF in the considered band). We found only one resolved substructure (Section \ref{subsec:substructure}). Following \cite{Zanella2019}, we used the fitting algorithm \texttt{GALFITM} \citep{Vika2013} that allows to simultaneously fit multiple images of the same galaxy taken at different wavelengths. After the subtraction of the best-fit model, we visually inspected all the residuals to verify the reliability of the fits. We show the results of our fitting procedure in Appendix \ref{app:morphology}.

At the redshift of our targets ($z = 6$), the \textit{HST}/WFC3 $z'$ bandpass includes the Ly$\alpha$ emission, while the \textit{JWST}/NIRCam F444W bandpass includes the H$\alpha$ line. The UV data produced by \code{skirt} do not include line emission (see Section \ref{subsec:emissionmodel}), thus no line subtraction is needed to obtain clean continuum maps. Similarly, this reasoning applies to the ALMA continuum and \cii~emission, and therefore the fluxes estimated with the described procedure do not need further corrections.

\subsection{Size measurements}
\label{subsec:size_measurements}

The galaxy disks are fitted with a S\'ersic profile and therefore we can also measure their effective radius. In the case of isolated galaxies the disks are barely resolved in the optical \textit{HST} bands and unresolved in the infrared \textit{JWST} images that have coarser resolution. They are always resolved in the \cii~maps (Section \ref{subsec:integrated}). When fitting the images, \texttt{GALFITM} convolves the S\'ersic model by an input PSF and returns the deconvolved effective radius. For this reason some of the effective radii reported in Table \ref{tab:fluxes_sizes} are smaller than the angular resolution of the mock observations. Since for the same galaxy \texttt{GALFITM} estimates comparable disk radii -- irrespective of the angular resolution of the observation -- the deconvolution procedure can be seen as reliable. To further check whether the limited sensitivity of our mock observations strongly affects the size estimates, we ran \texttt{GALFITM} on noiseless maps. Given the depth of our mock observations, we do not find systematic differences and in all bands the measured effective radii are consistent within the uncertainties with those estimated from the noisy maps.

When performing the fits we let \texttt{GALFITM} free to vary the S\'ersic index of the disks. We find that the clumpy disk has S\'ersic index $n \sim 1.5$ in all bands. The merger instead has a S\'ersic index $n \sim 1$ in the optical bands, $n \sim 2$ in the infrared bands, and $n \sim 0.5$ in the \cii~map. These results do not depend on the angular resolution of our images. Often in the literature, to limit the number of free parameters especially in the case of low signal-to-noise ratio (S/N), an exponential disk ($n = 1$) is fit. We checked that if we fit the disks adopting exponential profiles, the effective radii remain consistent with the case of a free S\'ersic model, within the uncertainties (we find changes $\lesssim 30$\% and no systematic trends with the observing band, Figure \ref{fig:profiles}). Finally, we measured the \cii~effective radius by fitting the line visibilities with the \code{CASA} task \texttt{UVMODELFIT}, adopting an exponential profile, following the procedure described by \cite{Fujimoto2020}. We obtained radii that are fully consistent with the ones measured with \code{GALFIT} in the image plane. Throughout the rest of the paper we report the \code{GALFIT} measurements, so to adopt consistent methods for optical, infrared, and submillimeter datasets.

\begin{figure}
\centering
\includegraphics[width=0.45\textwidth]{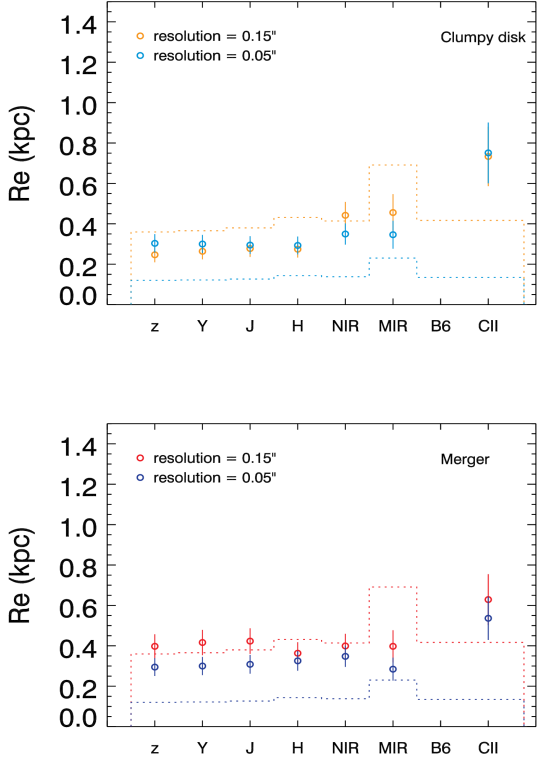}
\caption{Effective radius of the galaxy disk in different observing bands and with
  different spatial resolution. \textbf{Top:} Clumpy
  galaxy case. \textbf{Bottom:} merger case. The radius measured in the mock observations with $\sim 0.15$'' (yellow and red open circles for the clumpy galaxy and merger case respectively), and $\sim 0.05$'' (cyan and blue open circles for the clumpy galaxy and merger case respectively) resolution are reported. The dashed lines show the effective resolution (PSF's FWHM/2) of the mock observations in each band.}
\label{fig:radii}
\end{figure}

\begin{figure*}
\centering
\includegraphics[width=\textwidth]{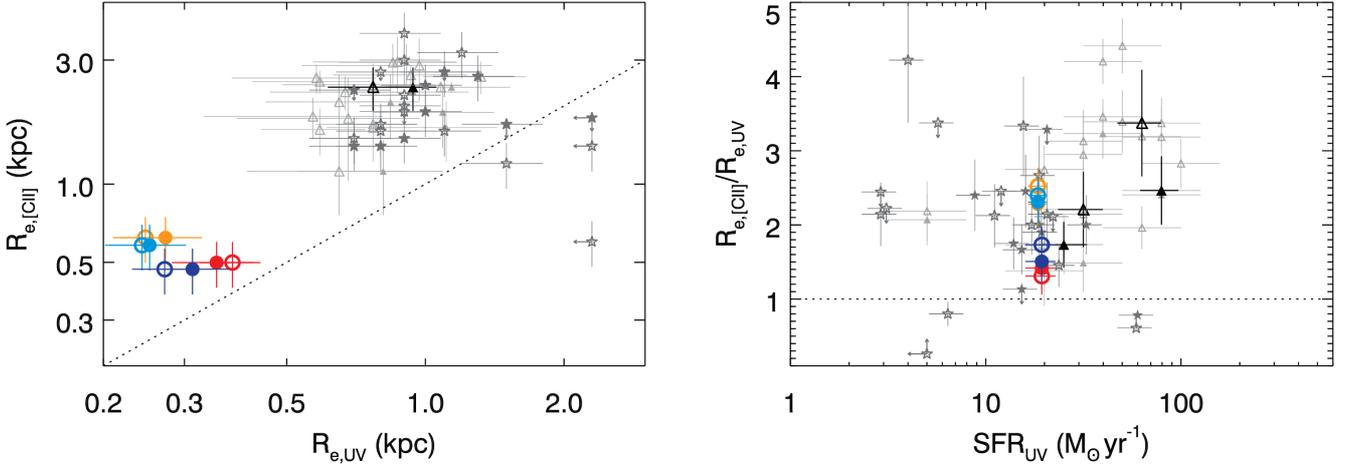}
\caption{Physical sizes of mock and observed galaxies. \textbf{Left}: comparison of the [CII] and rest-frame UV circularized effective radii. Literature results from the ALPINE survey \citep{Fujimoto2020} are reported in gray, with open and filled triangles indicating radii measured in the F814W and F160W bands respectively. The black empty and filled triangles indicate the median of the literature data (for the F814W and F160W bands) and their error bars represent the standard error of the median. Literature results from the collection of \citet[including data from \citealt{Capak2015}, \citealt{Barisic2017}, \citealt{Faisst2017}, \citealt{Willott2015}, \citealt{Jones2017}, \citealt{Ouchi2013}, \citealt{Matthee2017}, \citealt{Pentericci2016}, \citealt{Smit2018}, \citealt{Maiolino2015}, \citealt{Ota2014}, \citealt{Carniani2017}, \citealt{Inoue2016}]{Carniani2018} are shown as gray stars. Empty stars indicate multi-component systems, whereas filled stars indicate individual galaxies. Our measurements for the clumpy galaxy (orange and cyan circles) and merger (red and blue circles) are reported. The orange and red circles indicate a spatial resolution of $\sim 0.15$'', whereas the cyan and blue circles represent the lensed case (resolution $\sim 0.05$''). We report the radii measured on the rest-frame ($z'$) and optical ($H$) bands (empty and filled circles), to compare with the literature. The equality R$_\mathrm{e,[CII]} = $R$_\mathrm{e,UV}$ is shown (dotted black line). \textbf{Right}: [CII]-to-UV circularized effective radii ratio as a funcion of UV SFR. The black stars indicate the median ratios in bins of SFR. Colors and symbols are the same as in the left panel.}
\label{fig:fujimoto}
\end{figure*}

\subsection{Estimate of the flux and size uncertainties}
\label{subsec:uncertainties_flux}

To estimate the uncertainties associated to our flux and size measurements we performed 1000 Monte Carlo simulations. We injected a fake PSF or S\'ersic profile at the time on top of each map. The structural parameters of these components were randomly chosen in the range spanned by the substructures and disks in our mock observations. We then treated these images with the same procedure detailed in Section \ref{subsec:flux_measurements}. To determine the uncertainties associated to the flux of the substructures, we divided the simulated PSFs (or S\'ersic profiles) in bins based on the contrast between their luminosity and that of the underlying disk, at the location of the substructure. This was a necessary step because the accuracy of \texttt{GALFITM} in estimating the flux largely depends on the contrast with respect to the underlying disk. For each injected PSF we computed the difference between the known input flux and the one retrieved by \texttt{GALFITM}, in each contrast bin. The standard deviation of the sigma clipped distribution of these differences gave us the flux uncertainties.

Given the fluxes estimated by \texttt{GALFITM} and the associated uncertainties, we determined the S/N and considered as detections only the substructures with S/N $\gtrsim 3$. If in a given band we obtained a S/N $< 3$, we calculated a 3$\sigma$ upper limit based on the estimated uncertainty. Using 3$\sigma$ upper limits when studying \textit{HST} data is standard in the literature. For ALMA data some works use 3$\sigma$ upper limits, whereas others prefer more conservative 5$\sigma$ limits due to the correlated noise of interferometric observations. In the rest of this work we consider 3$\sigma$ limits also for the sub-millimeter bands, as most of the substructures detected in ALMA are also detected in at least one optical band. All but one structure would still be detected if we were to consider 5$\sigma$ limits instead and our results would not change. 

The uncertainties on the effective radii of the S\'ersic components were also derived by considering Monte Carlo simulations, with the same procedure adopted for flux uncertainties.

\section{Results}
\label{sec:results}

In this Section we discuss the structural properties (i.e. effective radius, S\'ersic index) of the clumpy disk and merger, and compare them with observational results. Furthermore, we investigate the presence and detectability of disks substructures, depending on spatial resolution and spectral band.
 
\subsection{Spatially-integrated galaxy properties}
\label{subsec:integrated}

Our galaxies are clearly detected (S/N $\gtrsim 5$) in the optical and infrared bands, as well as in the \cii~emission line maps, irrespective of the resolution of the observations. Instead, they are undetectable in the sub-millimeter continuum maps (e.g. Band 6). This is consistent with literature results, where the continuum at $\sim$ 158 $\mu$m is often undetected, even when the \cii~ is observed with high S/N (\citealt{Capak2015}, \citealt{Tamura2019}, \citealt{Bakx2020}).

We fitted the 2D light profile of the clumpy disk and merger using a S\'ersic profile (Section \ref{subsec:size_measurements}) and determined the effective radius of the disks. We find that they both have $R_\mathrm{e} \sim 300$ pc in the optical and infrared bands (Figure \ref{fig:radii}). The galaxies are marginally resolved or unresolved (especially in the \textit{JWST}/MIRI band) in the low-resolution case, whereas they are resolved at high-resolution. The disk size estimate does not vary, within the uncertainties, when changing the angular resolution of the observations, indicating that the fits are robust when the sources are detected with S/N $\gtrsim 5$ at angular resolution $< 0.15"$. The effective radius measured from the \cii~maps instead is systematically $\sim 1.5 - 2.5$ times larger than the optical one and ranges between $600$ pc (for the merger) and $700$ pc (for the clumpy disk).

The fact that \cii~sizes are systematically larger than the optical ones has been already reported (\citealt{Carniani2017}, \citealt{Fujimoto2019}, \citealt{Ginolfi2020}). Recently the ALPINE survey observed a sample of $z = 4 - 6$ galaxies and found, on a statistical basis, that the \cii~ sizes are $\sim 2 - 3$ times larger than the rest-frame UV sizes measured in the $z$ band, and $\sim 1.5 - 2$ times larger than the rest-frame optical sizes measured in the $H$ band\footnote{The \cii~observations of the ALPINE survey have been performed with a spatial resolution of $\sim 0.7"$ \citep{Lefevre2019}. To understand whether the higher spatial resolution of our mock maps was biasing the comparison with the ALPINE results, we have created a set of \cii~ maps with $\sim 0.7$'' angular resolution. The \cii~effective radius that we retrieved is $800 \pm 160$ pc, slightly larger than the one measured at higher resolution, but consistent within the uncertainties. We therefore conclude that the high angular resolution of the mock observations does not bias our comparison.} \citep{Fujimoto2020}. In Figure \ref{fig:fujimoto} (left panel) we compare the sizes measured in the $z$, $H$, and \cii~maps of our galaxies with the results from \cite{Fujimoto2020}. We find that the \cii~is systematically more extended than the rest-frame UV emitting regions, with \cii -to-UV size ratios similar to those reported in the literature. We also find that the R$_\mathrm{e,[CII]}$/R$_\mathrm{e,z}$ ratio is $\sim 20 - 30$\% smaller than the R$_\mathrm{e,[CII]}$/R$_\mathrm{e,H}$ ratio, in agreement with literature results. 

Our galaxies however seem to be more compact than those observed by \cite{Fujimoto2020}, having both UV and \cii~effective radii a factor $\sim 2.5$ smaller than those reported in the literature. 
This might be due to the fact that the two stages of {Alth\ae a} that we have analyzed were originally found at higher redshift ($z \gtrsim 7$, see Section \ref{sec:mock_obs}), where galaxies are expected to have smaller sizes, at fixed stellar mass (\citealt{Allen2017}, \citealt{Whitney2019}). More simulations are needed to understand the origin of this potential discrepancy with observations.

We further investigated whether the ratio of the \cii -to-UV radii could be related to the galaxy SFR (Figure \ref{fig:fujimoto}, right panel). Here we consider the unobscured SFR as derived from the UV luminosity, without correction for dust extinction. We complemented the literature sample of \cite{Fujimoto2020} with that of \cite{Carniani2018}, and compared with our mock observations. A trend of R$_{\mathrm{e,\cii}}$/R$_{\mathrm{UV}}$ with SFR is observed when considering the galaxies from \cite{Fujimoto2020} only, but it is washed out when adding the sample from \cite{Carniani2018}. We notice that while \cite{Fujimoto2020} excludes disturbed systems from their sample, a large fraction of the sources reported by \cite{Carniani2018} can be considered as mergers (Section \ref{sec:discussion}). This could be a reason why the two samples show discrepant results about the existence of a R$_\mathrm{e,[CII]}$/R$_\mathrm{e,UV}$ trend with SFR. This seems to be consistent with our mock observations showing that the clumpy disk and merger, despite having similar SFRs show significantly different ratios (R$_{\mathrm{e,\cii}}$/R$_{\mathrm{UV}} \sim 2.5$ for the clumpy disk and $\sim 1.5$ for the merger). More data on both the observational and theoretical side are needed to clarify whether the UV-to-\cii~ratio indeed scales with the galaxy SFR.

We highlight that \cite{Fujimoto2020} reports circularized effective radii (R$_\mathrm{e,circ} = $R$_\mathrm{e}\sqrt{q}$, where $q$ is the axis ratio). To perform a consistent comparison, in Figure \ref{fig:fujimoto} we also report the circularized effective radii of our galaxies, whereas in all other figures we show R$_\mathrm{e}$. Given that the axis ratio of our galaxies is $q \gtrsim$ 0.7, the difference between R$_\mathrm{e}$ and R$_\mathrm{e,circ}$ is $\lesssim 20$\%. 
\subsection{Galaxy substructure}
\label{subsec:substructure}

Our simulations show that galaxies in the early Universe do not appear simply as smooth disks, but they rather have complex substructures such as star-forming clumps, merging satellites, proto-spiral arms, and rings around the nucleus (Figure \ref{fig:maps16}, \ref{fig:maps23}). From our mock observations we can assess whether these substructures are detectable or are lost due to the limited spatial resolution and sensitivity. It is also possible to identify the optimal observational bands for substructure detection, and compare data taken at different wavelengths.

In the following we denote \quotes{clumps} as \textit{star-forming regions arising from disk gravitational instabilities}, and \quotes{satellites} as \textit{small galaxies of external origin merging with \althaea 's disk}. Note that satellites are embedded in their dark matter halo, whereas clumps are not \citep{Kohandel2019}. Finally, clumps detected in our mock maps are typically in virial equilibrium \citep{leung:2019}, and are therefore stable, self-gravitating structures.

We investigated whether the extended \cii~could be emitted by unresolved, dust-obscured satellites or clumps by comparing the effective radii estimated from mock observations with different spatial resolution. The \cii~and UV sizes that we measured are comparable in the low- and high-resolution cases (Figure \ref{fig:radii}), despite the presence of clumps that can only be resolved when the spatial resolution is $\sim 0.05$" or better (see Section \ref{subsubsec:resolution}). Furthermore, our clumpy disk shows two clumps aligned along the galaxy major axis (Figure \ref{fig:maps16_0.05}). They are below the detection threshold both in \cii~and UV maps, and they are only detected (and therefore deblended) at near- and mid-infrared bands. Nevertheless, the effective radius of the disk measured in the \textit{HST} optical and \textit{JWST} infrared bands is comparable within the uncertainties. This suggests that the presence of unresolved satellites and/or clumps does not substantially bias the measurement of the disk effective radius. Given the fact that we estimated the \cii~size with the same procedure as the optical one, we can as well conclude that the \cii~extended emission is not biased by the presence of undetected substructures. 

We conclude that at $z \sim 6$ the \cii~emission is intrinsically more extended than the optical one likely due to the joint effect of the carbon photoionization produced by UV photons emitted by the galaxy itself and penetrating in the surrounding neutral medium, and by the outflows produced by supernovae and massive stars that expel \cii~outside the disk. 

\subsubsection{Spatial resolution}
\label{subsubsec:resolution}

\begin{figure}
\centering
\includegraphics[width=0.45\textwidth]{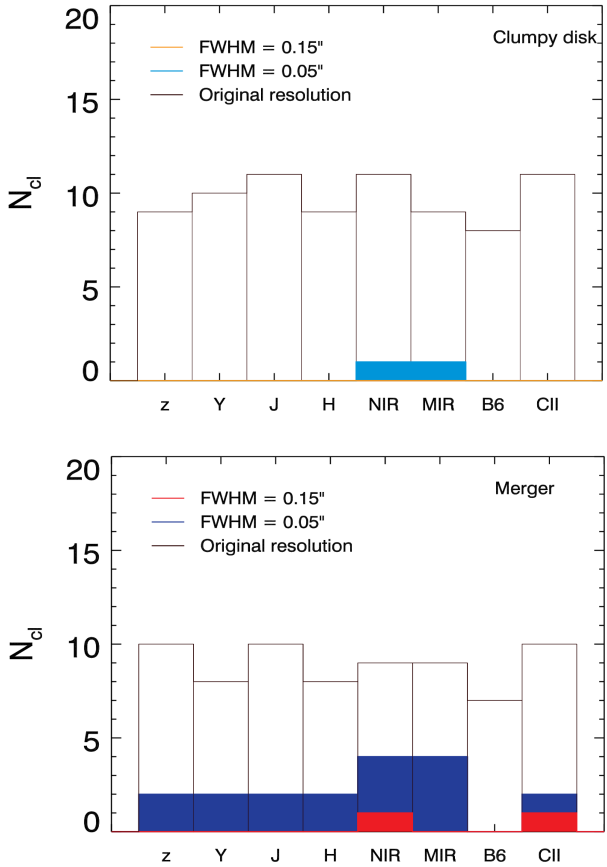}
\caption{Number of substructures found in different observing bands and with
  different spatial resolution. \textbf{Top:} clumpy
  galaxy case. \textbf{Bottom:} merger case. The number of
  substructures found in the simulations with original resolution (black empty histogram),  mock observations with $\sim 0.15$'' (red filled histogram), and
$\sim 0.05$'' (cyan and blue filled histogram) resolution are reported.}
\label{fig:nclumps}
\end{figure}

We determined the number of clumps and satellites detected in the maps with the original resolution of the simulation ($\sim 25$ pc) by using \code{SExtractor} (\citealt{Bertin1996}, Section \ref{sec:analysis}). Both our clumpy disk and merger show a similar number of substructures (Figure \ref{fig:nclumps}), between 7 and 11 depending on the observing band. When we consider the low- and high-resolution galaxy maps (resolution $\sim 0.15$'' and $\sim 0.05"$ respectively), the number of detected substructures drastically decreases. The clumps that are closer to the galaxy nucleus are blended and cannot be detected against the galaxy disk. The fainter and smaller clumps and satellites are instead undetected due to the lack of sensitivity and resolution of the mock observations. Only the substructures with high enough contrast against the galaxy disk or the background (e.g. those that are further away from the nucleus and/or brighter) can be detected. 

We do not detect substructures in the low-resolution clumpy galaxy case, which appears as a smooth S\'ersic disk (Figure \ref{fig:maps16_0.15}). In the merger case we only detect one substructure in the near-infrared \textit{JWST}/NIRCam band and in the \cii~map, but it does not appear at optical wavelengths. This is a merging satellite with stellar mass M$_\star = 1.3 \times 10^9$ M$_\odot$, molecular gas mass M$_\mathrm{H2} = 0.4 \times 10^7$ M$_\odot$, and SFR $= 10.8$ M$_\odot$ yr$^{-1}$. Its projected galactocentric distance  ($\sim 2.5$ kpc), and high contrast allowed us to deblend it from the disk and detect it.

When considering the high-resolution case instead (resolution $\sim 0.05$'') more substructures appear. The clumpy disk shows one clump, detected in the near- and mid-infrared bands. The merger shows two substructures in the optical and \cii~maps, and five at near- and mid-infrared wavelengths. The stellar mass of these substructures ranges between $M_\star \simeq (0.3 - 2.5) \times 10^8 M_\odot$, their total gas mass $M_\mathrm{gas} \sim (1 - 4) \times 10^7 M_\odot$, and their SFR $\simeq 0.2 - 1.0$ M$_\odot$ yr$^{-1}$. The most massive one and the most distant from the galaxy nucleus is a satellite, whereas the innermost ones are clumps. We note however that even in our high-resolution cases, only the clumps that are laying in the outskirts of the disk are detectable, whereas the innermost ones have a too low contrast with the disk to be deblended and studied. We note however that the detection of clumps might be easier when targeting galaxies with larger effective radii than {Alth\ae a}, as the contrast of the substructures with the galaxy disk is higher.

\subsubsection{Observing bands}
\label{subsubsec:bands}

Finally, we investigate how the band used to carry out the observations affects the detection of galaxies substructure by comparing the number of clumps and satellites found in maps at different wavelengths. It seems that the near- and mid-infrared \textit{JWST} maps are the most suitable ones for this study, as they show the highest number of detected substructures (Figure \ref{fig:nclumps}). 

\begin{figure*}
\centering
\includegraphics[width=\textwidth]{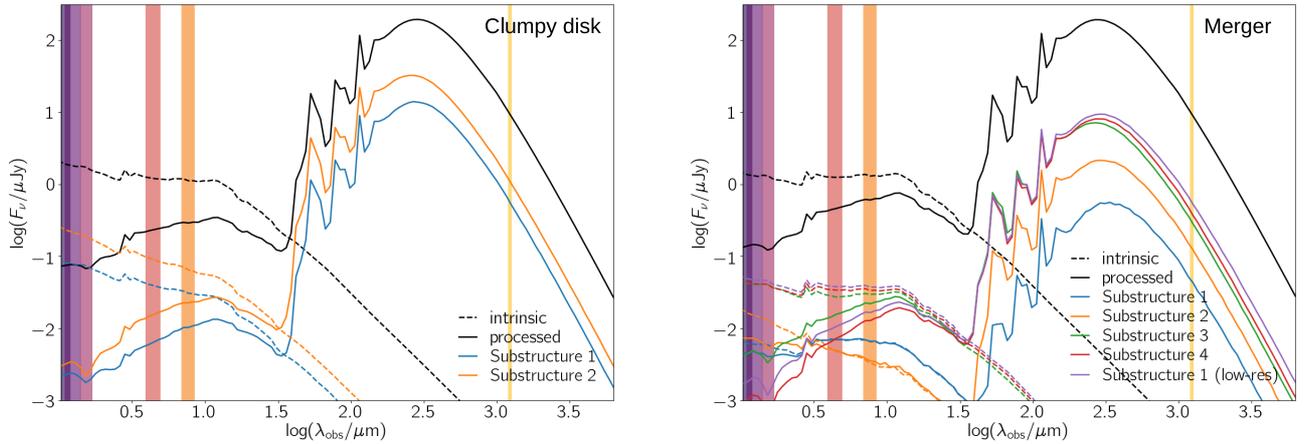}
\caption{Spectral energy distribution of the clumpy galaxy and merger. We show the SED of the integrated galaxy (black curve) and those of the individual substructures found in our mock maps (colored curves). We display the intrinsic emission (dashed curves) and the ``observed'' one that also includes the effects of dust extinction (solid curves). The vertical colored strips indicate the observing bands used to create our mock maps (\textit{HST/z', J, Y, H}, \textit{JWST/}NEAR-IR, MID-IR, ALMA/Band 6). \textbf{Left panel:} clumpy disk. \textbf{Right panel:} merger.}
\label{fig:sed}
\end{figure*}

This can be seen when considering the SED of the individual substructures found in our mock maps. In Figure \ref{fig:sed} we show that indeed those with the highest dust extinction (e.g. number 1 and 2 in the clumpy high-resolution case; number 3 and 4 in the merger high-resolution case) are undetected in the \textit{HST} bands (Table \ref{tab:fluxes_sizes}). At optical wavelengths we could only detect the unobscured substructures (e.g. number 1 and 2 in the merger, high-resolution case) which are also metal-poor and/or gas-poor and, due to these reasons, are undetected in \cii \ (Table \ref{tab:physical_prop}). The \textit{JWST} bands instead allow us to simultaneously detect both unobscured (metal-poor) and dust-obscured (metal-rich) substructures, being therefore ideal for the study of galaxy clumps and satellites.

The number of substructures detected at optical and sub-millimeter wavelengths is similar, although those found in the \textit{HST} and ALMA \cii~maps are not co-spatial (see e.g. Figure \ref{fig:maps23_0.05}). The morphology of the clumpy disk and merger appearing at optical and sub-millimeter wavelengths is the following:

\begin{itemize}
    \item \textbf{Clumpy disk.} The galaxy disk is detected in both \textit{HST} optical and ALMA \cii~maps and the emissions are co-spatial. We do not detected substructures at these wavelengths (the only substructures are detected in the infrared \textit{JWST} bands), irrespective of the spatial resolution (Figure \ref{fig:maps16_0.05} and \ref{fig:maps16_0.15}).
    \item \textbf{Merger.} Also in this case, the galaxy disk is detected both in optical and \cii~bands, although a more complex morphology is observed. In the low-resolution case, only the galaxy disk is visible at optical wavelengths, whereas the \cii~map also shows an additional component. It is a small satellite (stellar mass $M_\star = 1.3 \times 10^9 M_\odot$, total gas mass $M_\mathrm{gas} = 3.4 \times 10^9 M_\odot$) merging with the main galaxy (Figure \ref{fig:maps23_0.15}). It has a distance of $\sim 2.5$ kpc and it is unresolved at the resolution of these observations (0.15", corresponding to $\sim 0.8$ kpc at this redshift). \\
    More substructures are seen at $\sim 0.05$" resolution (Figure \ref{fig:maps23_0.05}). In particular, two clumps are detected at optical wavelengths (but not in \cii) and, viceversa, other two structures are detected in \cii~(but not in the optical bands). The two optical substructures have a distance from the galaxy center of 0.7 -- 0.8 kpc, whereas the two sub-mm substructures are found respectively at a distance of $\sim$ 0.8 kpc and $\sim$ 2.5 kpc from the galaxy center. The substructure found at 2.5 kpc from the galaxy disk is a satellite (the same detected also with spatial resolution $\sim 0.15"$), whereas the others are clumps.
\end{itemize}

The origin of the optical versus \cii~displacement in our mocks is the role played by dust extinction and metallicity, as already pointed out by \cite{Vallini2015}, \cite{Pallottini2017b}, and \cite{Katz2017}. The most dust-obscured structures are detected in \cii~but not in the optical bands and, viceversa, the most metal-poor ones are detected at rest-frame UV wavelengths but not in the sub-mm (Figure \ref{fig:sed}).

\begin{center}
\begin{table*}
    \caption{Physical parameters of the detected substructures.}
\resizebox{\textwidth}{!}{%
\begin{tabular}{ccccccccc}
\toprule
\midrule
Galaxy    & Resolution & Component  & $M_\star$   &  SFR       & $M_\mathrm{g}$ & $M_\mathrm{H2}$  & $Z$      & $A_V$ \\
          &            &            & $10^8\msun$ &  $\msunyr$ & $10^8\msun$    & $10^6\msun$      & $\zsun$  &  \\ 
(1)       & (2)        & (3)        & (4)         & (5)        & (6)            & (7)              & (8)      & (9) \\
\midrule
Clumpy & High-resolution & Substructure 1    & 1.3         & 4.2        & 0.2            & 1.1              & 0.2   & 2.2 \\ 
       &                 & Substructure 2    & 2.7         & 3.9        & 0.9            & 0.8              & 0.02  & 2.1 \\ 
Merger & Low-resolution  & Substructure 1   & 4.7         & 1.7        & 1.6            & 1.9              & 0.1   & 1.5 \\ 
       & High-resolution & Substructure 1    & 0.4         & 0.2        & 0.1            & 0                & 0.1   & 0.1 \\ 
       &                 & Substructure 2    & 0.3         & 0.2        & 0.4            & 0.5              & 0.02  & 0.1 \\ 
       &                 & Substructure 3    & 1.6         & 0.3        & 0.3            & 0                & 0.1   & 0.8 \\ 
       &                 & Substructure 4    & 2.5         & 1.0        & 0.3            & 0.5              & 0.2   & 1.9 \\ 
\bottomrule
\end{tabular}
}
\begin{minipage}{0.96\textwidth}
\textbf{Columns:} (1) Galaxy: clumpy disk or merger. (2) Resolution: low- ($\sim 0.15"$) or high-resolution ($\sim 0.05"$). (3) Component: substructure found in the mock maps. (4) Stellar mass. (5) Star formation rate. (6) Gas mass. (7) Molecular gas mass. (8) Metallicity. (9) Dust extinction.
\end{minipage}    
\label{tab:physical_prop}
\end{table*}
\end{center}

\section{Discussion}
\label{sec:discussion}

In recent years increasingly large samples of high-redshift galaxies observed at both optical and sub-millimeter wavelengths with relatively high spatial resolution ($\sim 0.15" - 0.5"$) have been reported (e.g. \citealt{Mallery2012}, \citealt{Carilli2013}, \citealt{Willott2015}, \citealt{Maiolino2015}, \citealt{Capak2015}, \citealt{Inoue2016}, \citealt{Pentericci2016}, \citealt{Knudsen2016}, \citealt{Carniani2017}, \citealt{Barisic2017}, \citealt{Bradac2017}, \citealt{Carniani2018}, \citealt{Fujimoto2020}). Some of them appear as isolated disks, whereas others show morphologies with multiple components. These studies have also shown that multiwavelength datasets are crucial to study these primordial systems, as various tracers (e.g. optical versus \cii) reveal different morphologies, with some substructures being detected only at certain wavelength. As a result, some of these multi-component systems have co-spatial optical and sub-millimeter emission, whereas others show spatial offsets. In order to understand what is the nature of the multi-component systems currently observed in the literature (i.e. accreting satellites vs star-forming clumps), we compare their properties (separation of the multiple components, their spatial extent, their luminosity at different wavelengths) with those of the star-forming clumps and accreting satellite detected in the stages of \althaea\, analyzed in Sections \ref{sec:analysis} and \ref{sec:results}.

\subsubsection*{Separation of multiple components}

We compared the structure of our clumpy disk and merger with the morphology reported in the literature for galaxies observed with \textit{HST} in the optical and ALMA in the sub-millimeter.

\begin{itemize}
    \item \textbf{Single-component systems.} The $z = 6.17$ galaxy (CLM1) observed by \citet{Willott2015} appears as an individual source with spatially coincident \cii~and optical emission. Other two similar cases have been observed by \citet{Smith2017} at $z = 6.85$ and $z = 6.81$, although the beam size of the ALMA observations was $1.1" \times 0.7"$, and data with better angular resolution would be needed to confirm these results. In the sample collected by \citet{Carniani2018}, $\sim 60$\% of the sources have a single-component morphology and they all show spatially-coincident optical and \cii~emission. Finally, both \citet{Knudsen2016} and \citet{Bradac2017} report the discovery of lensed galaxies (magnification $\mu = 11.4 \pm 1.9$ and $\mu = 5.0 \pm 0.3$ respectively) showing spatially coincident \cii~and $Y$ band emission. No substructures are found in these galaxies, despite lensing allows to reach higher spatial resolutions.
    \item \textbf{Multi-component systems.} Among the $z = 5 - 7$ sources presented by \citet{Carniani2018} observed both with \textit{HST} and ALMA, $\sim 40$\% show multi-component morphologies. The substructures are separated by $\gtrsim 2$ kpc. For half of these targets the optical and sub-millimeter emissions are not co-spatial, with some components visible at optical wavelengths, whereas others detected in \cii.
\end{itemize}

The literature galaxies showing a single component are comparable to our clumpy disk case. The comparison with our simulations suggests that they are likely isolated, undisturbed disk galaxies, with a rather homogeneous distribution of dust and metals, allowing the detection of both optical and \cii~emission. Only the galaxy disk is detected in these cases and no substructures are observed.

The comparison with our mock observations suggests instead that the multi-component systems are likely galaxies undergoing mergers, rather than star-forming clumps formed \textit{in-situ} in the galaxy disk. The substructures reported in the literature in fact have offsets $\gtrsim 2$ kpc \citep{Carniani2018}, consistent with the distance of the satellite from the galaxy disk ($\sim 2.5$ kpc) in our merger case. Some of the satellites might be detected only in the optical or \cii, giving rise to the observed spatial offsets, due to different metallicity or dust content, as it is the case for \althaea~(Section \ref{subsubsec:bands}).

We point out that the galaxy disk in our mock observations is always detected both in the optical and sub-millimeter bands, so we do not have extreme cases of completely displaced optical and \cii~emission as the one reported by, e.g. \citet{Maiolino2015,Carniani2017}. While strong differences in the dust and metallicity content of the subcomponents in this system could explain the observed offsets, strong feedback cleaning the most vigorous star-forming regions might also play a role in this case, as suggested by \cite{Maiolino2015} and \cite{Gallerani2018}. More simulations of high-redshift galaxies are required to assemble a statistically significant sample, investigate whether simulated systems with completely offset optical and sub-millimeter emissions exist, and understand what is the role played by feedback in the assembly of early galaxies.

\subsubsection*{Spatial extent}

Current observations reporting multiple substructures have been performed at $z \sim 5 - 7$ with angular resolution $\gtrsim 0.15"$. Most of these substructures are spatially resolved and have rest-frame UV sizes R$_\mathrm{e,UV} \gtrsim 0.7$ kpc and infrared sizes R$_\mathrm{e,\cii} \gtrsim 1$ kpc \citep{Carniani2018}. Their sizes are comparable to those measured for individual, isolated galaxies (Figure \ref{fig:fujimoto}). This suggests that the multi-component systems currently reported in the literature are likely merging galaxies rather than disks hosting star-forming clumps formed \textit{in-situ}. Additionally, in Section \ref{subsubsec:resolution} we have shown that spatial resolution is key to study galaxies' substructure. \althaea's star-forming clumps are detected only in the high-resolution maps (resolution $\sim$ 0.05"), whereas in the low-resolution case only the galaxy disk and a satellite are visible. This indicates that higher resolution than currently achieved in observations is needed to detect the internal structure of galaxies. 

\subsubsection*{UV and \cii~luminosity}

Finally, we compared the optical and \cii~luminosity of single- and multi-component systems from the literature, with the disk and substructures identified in the mock maps of our clumpy disk and merger. In local galaxies a tight relation between the \cii~and rest-frame UV luminosity (or equivalently the unobscured star formation rate, SFR$_\mathrm{UV}$) has been observed (e.g. \citealt{Pineda2013}, \citealt{deLooze2014}, \citealt{Kapala2015}, \citealt{Herrera-Camus2015}). Recently it has been investigated whether the same relation holds also at high redshift ($z > 5$). 

Several studies have reported that high-redshift galaxies seem to be more scattered and often \cii --deficient with respect to local sources (\citealt{Willott2015, Pentericci2016, Bradac2017,Carniani2017, Harikane2018, Harikane2019}; for a theoretical interpretation see \citet{ferrara:2019} and \citealt{pallottini:2019}). However most of these studies did not consider the multi-component nature of high-redshift galaxies and reported them as single sources in the \cii~luminosity -- SFR plane. \cite{Carniani2018,Carniani2020} showed that when associating each \cii~subcomponent with its proper UV counterpart (when detected), the high-redshift sources follow on average the local L$_\mathrm{[CII]}$ -- SFR relation. They estimate the dispersion of the high-redshift relation to be 1.8 times larger than the one reported by \citet{deLooze2014} for local galaxies, but no systematic offsets are observed. As discussed by \citet{Carniani2018}, the larger dispersion at high-redshift might be explained by the presence of multi-component and complex systems in different evolutionary stages that are not common in the local Universe. Similar results have been recently found also by \citet{Matthee2019} and \citet{Schaerer2020}.

\begin{figure*}
\centering
\includegraphics[width=0.8\textwidth]{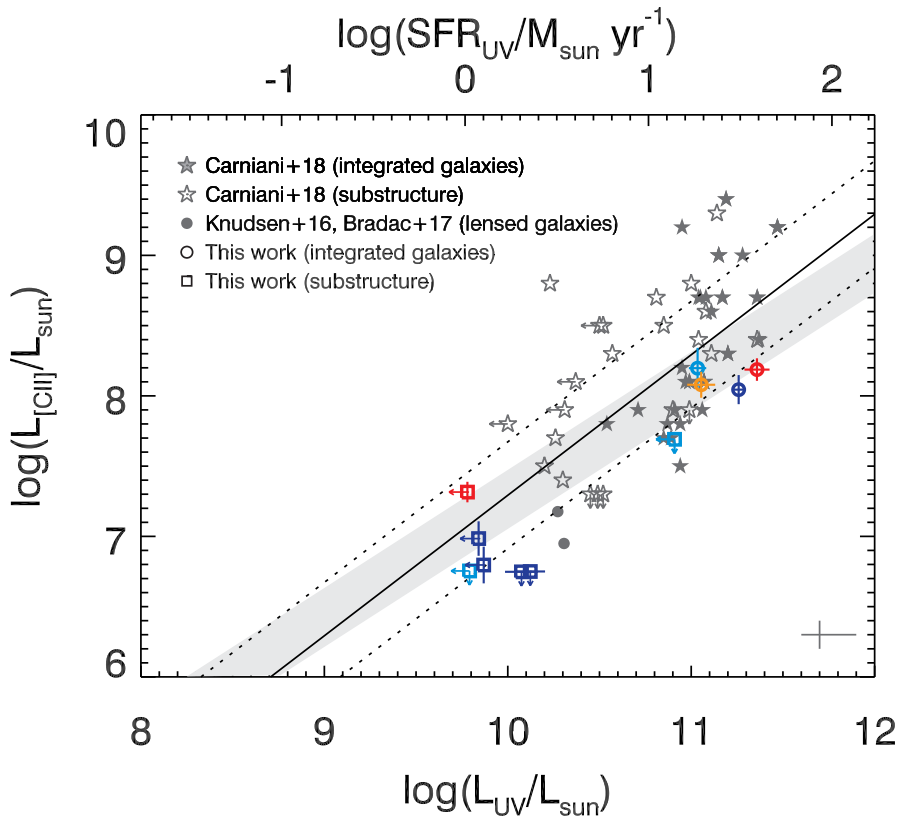}
\caption{Comparison of the \cii~and UV luminosity of disks and substructures. Literature results for non-lensed targets \citep[and references therein]{Carniani2018} are reported as gray empty and filled stars to indicate respectively individual substructures and the total luminosity of the system (given by the sum of the luminosity of all related substructures). Literature results for lensed targets \citep{Knudsen2016,Bradac2017} are reported as gray filled circles. Typical error bars are shown in the bottom right corner. Measurements and upper limits for our substructures (colored squares) and disks (colored circles) are shown (the colors are the same as in previous Figures). We report the relation between the [CII] luminosity and the unobscured SFR from \citealt{deLooze2014} (black line) and its 1$\sigma$ standard deviation (dotted black lines). The gray filled area represents the relation between SFR (from optical SED fitting) and \cii~luminosity for the ALPINE galaxy sample \citep{Schaerer2020}.}
\label{fig:uv_cii}
\end{figure*}

We have determined the sub-components location in the L$_\mathrm{[CII]}$ -- L$_\mathrm{UV}$ (or equivalently SFR$_\mathrm{UV}$) plane (Figure \ref{fig:uv_cii}) and compared our findings with literature works also reporting the unobscured SFR \citep{deLooze2014,Carniani2018,Schaerer2020}. Our galaxy disks and sub-components (i.e. satellites and clumps) seem to lay on a consistent relation with respect to that reported in the literature for $z \sim 5 - 7$ sources, with clumps and the satellite typically being 10 times fainter than the disk.

Figure \ref{fig:uv_cii} also shows that most of the substructures detected in our mock observations have UV and \cii~luminosities that are below the detection threshold of current observations. We therefore conclude that most of the multi-component systems reported in the literature are galaxies undergoing major mergers, rather than small satellites or clumps that, as shown by our simulations, would have lower luminosity. This is also consistent with their relatively large sizes ($\gtrsim 0.15" \sim 1$ kpc) and typical separation ($\gtrsim 2.5$ kpc).

\subsection{Insights on the mass assembly of early galaxies}
\label{subsec:early_growth}

Detecting star-forming clumps embedded in the disk of early galaxies is key to determine the fraction of clumpy sources and the contribution of hierarchical merging and \textit{in-situ} secular growth to the mass assembly of early galaxies. \cite{Shibuya2016} analyzed a sample of Lyman Break Galaxies at $ z = 4 - 8$ with available multi-band \textit{HST} photometry and found $\sim 15\% - 20\%$ of their sample galaxies to show substructures at these early times. When complementing this dataset with observations of lower-redshift galaxies ($z \sim 0 - 3$) they find that the fraction of multi-component systems ($f_\mathrm{multi}$) reaches a peak at $z \sim 2$ ($f_\mathrm{multi} \sim 60\%$) and then declines again towards lower redshift ($f_\mathrm{multi} \sim 40\%$ at $z\sim 1$). They compared the redshift evolution of $f_\mathrm{multi}$ with the expected evolution of minor and major merger fractions \citep{Lotz2011}. They found that mergers cannot fully explain the observed trend of $f_\mathrm{multi}$ and that violent disk instability giving rise to \textit{in-situ} massive clumps seems to play a major role in galaxy mass assembly. The fact that the $f_\mathrm{multi}$ evolution with redshift seems to closely follow the star formation rate density evolution \citep{Madau2014} reinforces this scenario. 

However, the results by \cite{Shibuya2016} at $z \gtrsim 4$ are based on observations of a very specific galaxy population (i.e. Lyman Break Galaxies). Due to their sample selection and the lack of sub-millimeter data, some dust-obscured substructures might have not been detected and therefore the fraction of multi-component systems at these redshifts might be higher. In the sample by \cite{Carniani2018} that combines optical and \cii~observations, the fraction of multi-component systems is $\sim 40\%$ at $z \sim 5 - 7$ and most of them seem to be consistent with being major (or possibly minor) mergers (Section \ref{sec:discussion}). 

In the near future it will be therefore key to acquire multi-wavelength observations of statistical samples of high-redshift galaxies with exquisite spatial resolution to better constrain the evolution of $f_\mathrm{multi}$ with redshift, understand what are the properties and nature of individual substructures (i.e. star-forming clumps vs merging satellites), and clarify what are the mechanisms driving galaxy formation at early epochs.

With current facilities very high spatial resolution can be achieved by observing lensed sources (with magnification factor $\mu \gtrsim 10$). However assembling large samples of lensed $z \sim 5 - 7$ galaxies is challenging. Furthermore, lensed sources are preferentially low mass, compact galaxies, that are expected to be less clumpy than more massive, larger targets \citep{Bournaud2014, Shibuya2016, Guo2018}. Likely for this reason no substructures have been detected in the lensed $z \sim 6$ galaxies observed so far with both \textit{HST} and ALMA by \cite{Knudsen2016} and \cite{Bradac2017}. To date only \citet{Vanzella2019} reports one lensed target at $z = 6.143$ that shows internal substructure; the target is a highly magnified ($\mu \sim 20$) dwarf galaxy hosting an extremely dense star-forming region (size $< 13$ pc, stellar mass $< 10^6$ M$_\odot$); this young (age $< 10 - 100$ Myr), moderately dust-obscured (E(B-V) $< 0.15$) star cluster was identified in a deep \textit{HST} pointing and spectroscopically confirmed with adaptive optics-assisted MUSE observations. No submillimeter data showing the gas and dust content of this target are currently available, but they would be key to gain a complete picture of the internal structure of this galaxy.

With the next generation of telescopes (e.g. \textit{JWST}, ELT) and instruments (e.g. MAVIS, the new adaptive optics-assisted visible imager and spectrograph proposed for the Very Large Telescope) it will be possible to investigate the internal structure of non-lensed targets (e.g. an angular resolution of $\sim 0.03" - 0.05"$ will be achieved at $\lambda \sim 0.7 - 1.5 \mu$m with \textit{JWST}/NIRCam and $\sim 0.02$" in $V$ band with VLT/MAVIS), collecting statistical samples and understanding what is the role of secular evolution in the mass assembly and evolution of early galaxies. The future facilities operating at infrared wavelengths will also allow the community to detect dust-obscured clumps and satellites, finding the rest-frame optical counterpart of currently detected \cii~substructures (e.g. with the $\textit{JWST}$/NIRCam near- and mid-infrared filters). This is seen in our mock observations as well, where the dust-obscured satellite lacking a rest-frame UV counterpart is instead detected in the \textit{JWST} maps (Section \ref{subsec:substructure}). An additional piece of information will come from spatially-resolved metallicity measurements. By combining ALMA observations of the FIR [OIII]$52\mu$m (or [OIII]$88\mu$m) emission with \textit{JWST} optical hydrogen lines (e.g. H$\alpha$, H$\beta$, Pa$\alpha$) it will be possible to conduct spatially-resolved gas-phase metallicity measurements on sub-galactic scales \citep{Jones2020}. This will help distinguishing mergers from clumpy disks, get insights into gas mixing and feedback processes, and constrain the contribution of mergers and \textit{in-situ} growth to the early assembly of galaxies.

\section{Summary}
\label{sec:conclusions}

We have analyzed two stages of {Alth\ae a}, a typical $z \simeq 6$ Lyman Break galaxy found in the SERRA zoom-in cosmological simulation suite. In the first snapshot {Alth\ae a} appears as a clumpy disk, whereas in the second it is undergoing a merger with a small satellite (stellar mass ratio 1:8). We created mock optical ($z'$, $Y$, $J$, and $H$ \textit{HST}-like), infrared (NIRCam/F444W and MIRI/F770W \textit{JWST}-like) and sub-millimeter (Band 6 and \cii~ALMA-like) observations. We performed a 2D morphological analysis, considering maps with different angular resolutions (0.15" and 0.05"), and we deblended the emission of the galaxy disks from that of substructures (merging satellites or star-forming clumps). We found that:

\begin{itemize}
    \item Our mock galaxies show \cii~effective radii $\sim 1.5 - 2.5$ times larger than the optical ones. This is consistent with recent findings from the literature (e.g. \citealt{Carniani2018}, \citealt{Fujimoto2020}). We conclude that the observed \cii~halos arise from the joint effect of stellar outflows and carbon photoionization by the galaxy UV field, rather than from the emission of unresolved nearby satellites \citep{Gallerani2018, pizzati:2020}.
    \item With a spatial resolution of $\sim 0.15"$ we detect only one merging satellite at a distance of $\sim 2.5$ kpc from the galaxy nucleus. Star-forming clumps are instead embedded in the galaxy disk (distance $\lesssim 1$ kpc). We show that better resolution ($\sim 0.05"$) is required to detect these substructures at $z \sim 6$. 
    \item Star-forming clumps found in our mock observations follow the local L$_\mathrm{\cii}$ -- SFR$_\mathrm{UV}$ relation reported in the literature for galaxy disks, but sample the low-luminosity (L$_\mathrm{\cii} \lesssim 10^{7.5}$ L$_\odot$), low-SFR (SFR$_\mathrm{UV} \lesssim 3$ M$_\odot$ yr$^{-1}$) tail of the distribution.
    \item Only clumps with low dust extinction (A$_\mathrm{V} \simeq 0.1$) are detectable in the \textit{HST}-like UV bands, whereas the dust-obscured (A$_\mathrm{V} \sim 1$) and metal-rich ones are detected in \cii~maps. The \textit{JWST} bands seem to be the most suitable ones to detect substructures thanks to their simultaneous sensitivity to both low-metallicity and dust-obscured regions that are bright at infrared wavelengths.
    \item By comparing the spatial extent, UV and \cii~luminosity, and separation of the substructures found in the multi-components systems reported in the literature at $z \sim 5 - 7$ \citep{Carniani2018}, we conclude that current observations are likely detecting galaxies undergoing major mergers, rather than their internal star-forming clumps. Future telescopes (e.g. \textit{JWST}, ELT) and instruments (e.g. VLT/MAVIS) with better sensitivity and spatial resolution will allow us to study star-forming clumps in $z \sim 6$ galaxies and quantify their contribution to the mass assembly of early galaxies.
\end{itemize}

\section*{Acknowledgements}
We thank the referee, Nick Gnedin, for his comments that improved the clarity of the manuscript.
AF (PI), MK, AP, SC acknowledge support from the ERC Advanced Grant INTERSTELLAR H2020/740120. Any dissemination of results must indicate that it reflects only the author's view and that the Commission is not responsible for any use that may be made of the information it contains.
This research was partly supported by the Munich Institute for Astro- and Particle Physics (MIAPP) of the DFG cluster of excellence "Origin and Structure of the Universe".
Partial support from the Carl Friedrich von Siemens-Forschungspreis der Alexander von Humboldt-Stiftung Research Award (AF) is kindly acknowledged.
AZ acknowledges hospitality from Scuola Normale Superiore where part of this work has been developed.
We acknowledge the use of the Python programming language \citep{VanRossum1991}, Astropy \citep{astropy}, Cython \citep{behnel2010cython}, Jupyter \citep{LoizidesSchmidt2016}, Matplotlib \citep{Hunter2007}, NumPy \citep{VanDerWalt2011}, \code{Pymses} \citep{Labadens2012}, \code{pynbody} \citep{pynbody}, and SciPy \citep{scipy2019}.

\section*{DATA AVAILABILITY}
The data underlying this article were accessed from the computational resources available to the Cosmology Group at Scuola Normale Superiore, Pisa (IT). The derived data generated in this research will be shared on reasonable request to the corresponding author.




\bibliographystyle{style/mnras}
\bibliography{bibliography} 


\newpage
\appendix

\section{Morphological structure in the optical and submillimeter bands}
\label{app:morphology}

\begin{figure*}
\centering
\includegraphics[width=0.75\textwidth]{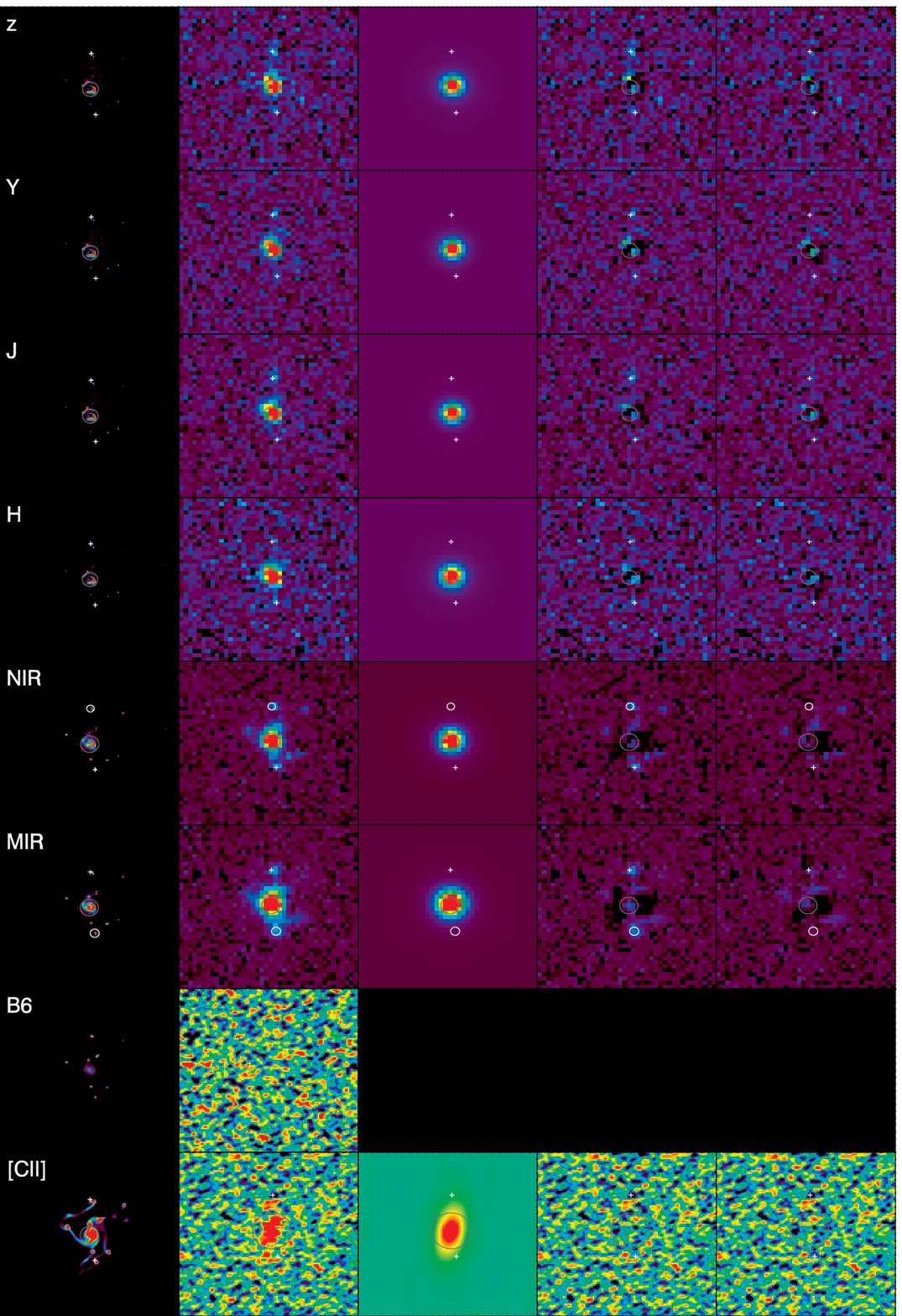}
\caption{Mock observations for the disk galaxy with resolution 0.05''. \textbf{From top to bottom:} \textit{HST}/ACS F850LP (\textit{z'}), \textit{HST}/WFC3 F105W (\textit{Y}, F125W (\textit{J}), F160W (\textit{H}); \textit{JWST}/NIRCam F444W (\textit{NIR}), \textit{JWST}/MIRI F770W (\textit{MIR}); ALMA Band 6 continuum and the \cii~pseudo-narrow band emission line map. \textbf{From left to right:} map with the original simulation resolution, mock maps with the resolution of observations ($\sim 0.15$''), \texttt{GALFITM} model for the diffuse component (single S\'ersic profile), residuals obtained subtracting the model (column 3) from the mock image (column 2), residuals obtained subtracting \texttt{GALFITM} best-fit model (including
the diffuse S\'ersic profile plus additional PSFs at the location of the substructures detected in the residuals shown in column 4) from the mock image (column 2). The dark gray circle indicates the center of the diffuse S\'ersic profile and its radius is equivalent to the disk effective radius. The white crosses and circles indicate the center of the substructures (respectively detected with S/N $>$ 3 or non-detected). The radius of the white circles is equivalent to the FWHM of the PSF (if they are unresolved) or to the effective radius of the best-fit S\'ersic profile (if they are resolved).
Each stamp has a size of 0.6'' $\times$ 0.6'' ($\sim 3.4 \times 3.4$ kpc
at $z \sim 6$), we adopt the same color bar in columns 2 to 4 and an inverse hyperbolic sine scaling.}
\label{fig:maps16_0.05}
\end{figure*}

\begin{figure*}
\centering
\includegraphics[width=0.75\textwidth]{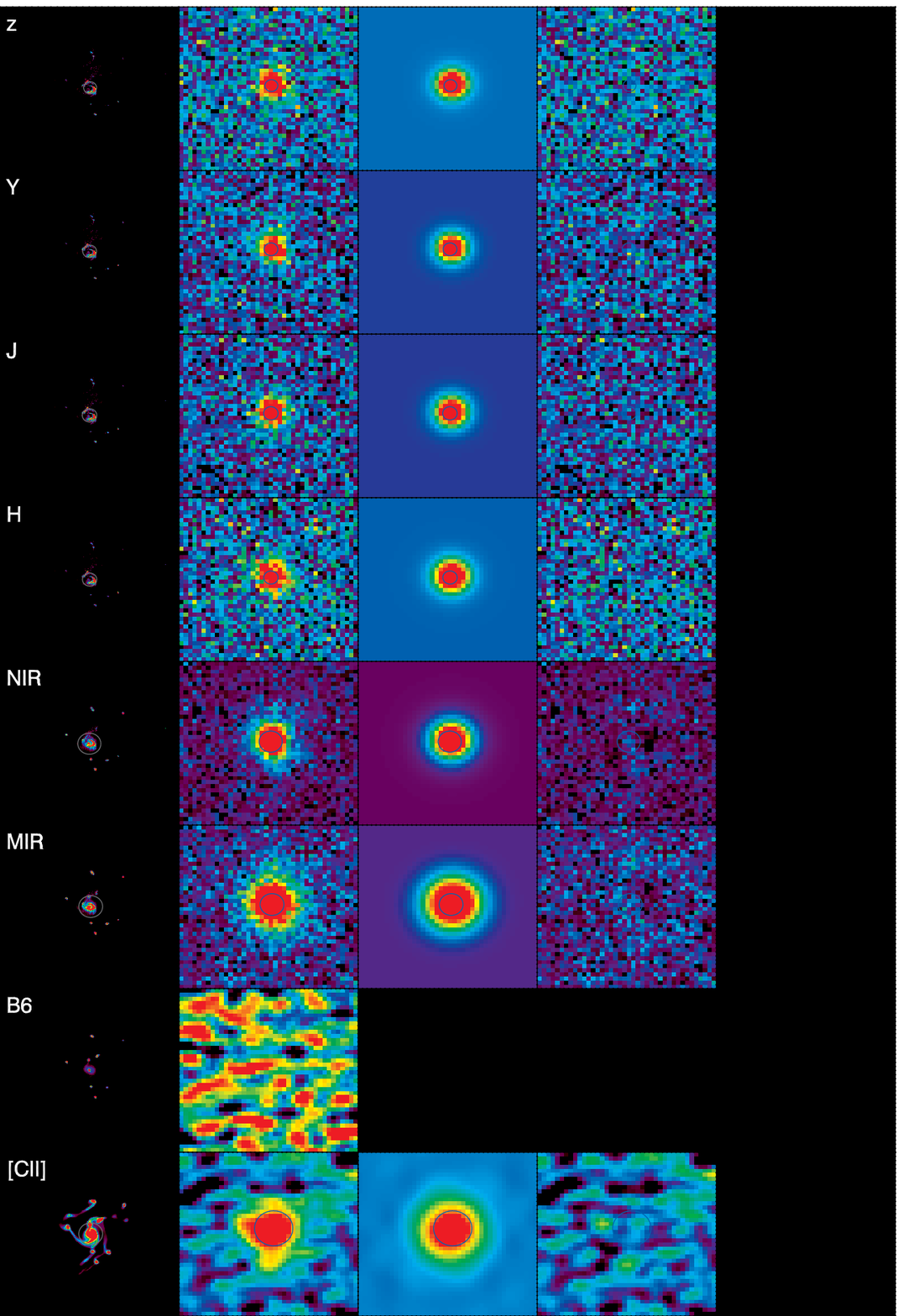}
\caption{Mock observations for the disk galaxy with resolution 0.05''. Images and symbols are the same as in Figure \ref{fig:maps16_0.05}}
\label{fig:maps16_0.15}
\end{figure*}

\begin{figure*}
\centering
\includegraphics[width=0.75\textwidth]{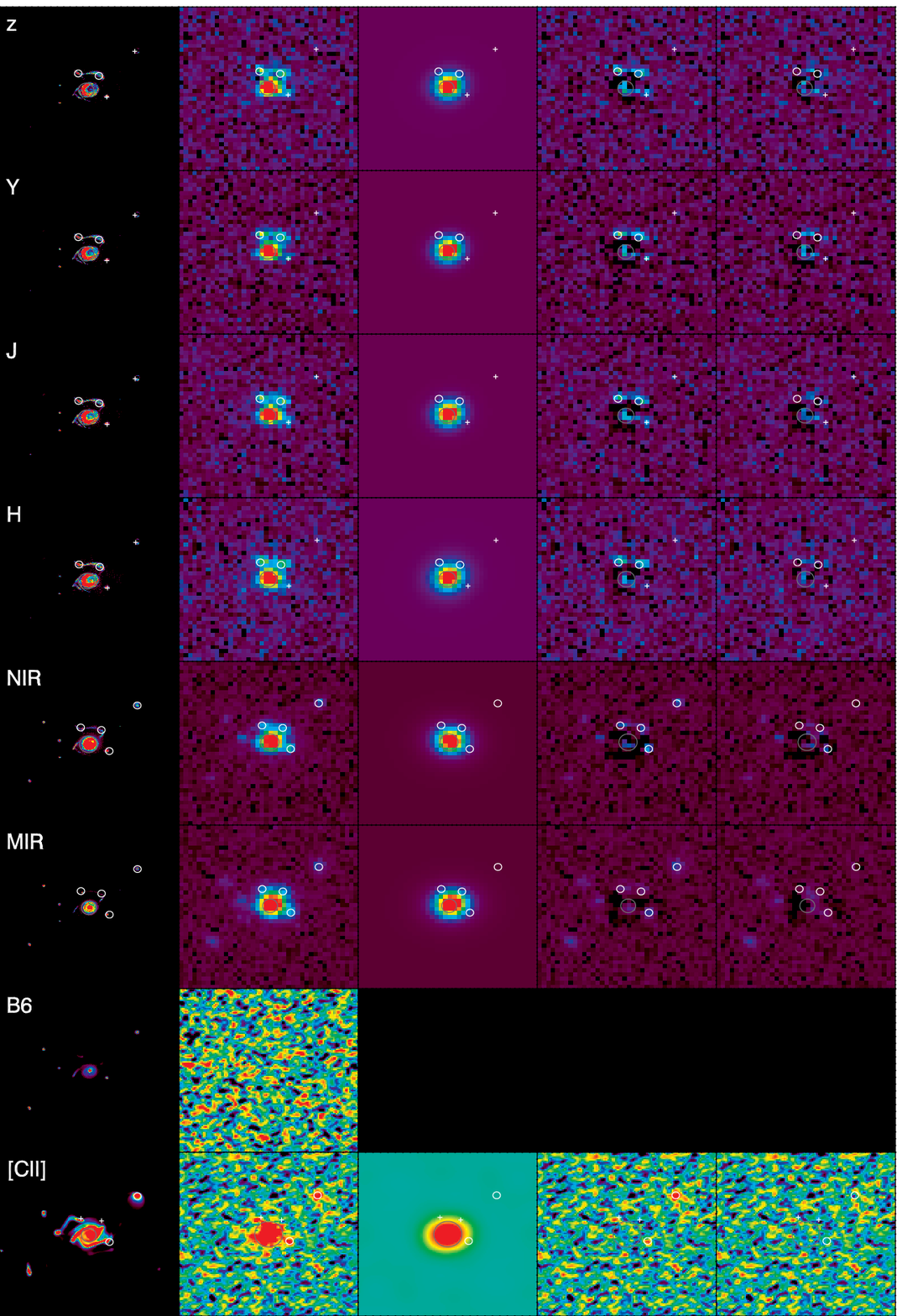}
\caption{Mock observations for the merger with resolution 0.05''. Images and symbols are the same as in Figure \ref{fig:maps16_0.05}}
\label{fig:maps23_0.05}
\end{figure*}

\begin{figure*}
\centering
\includegraphics[width=0.75\textwidth]{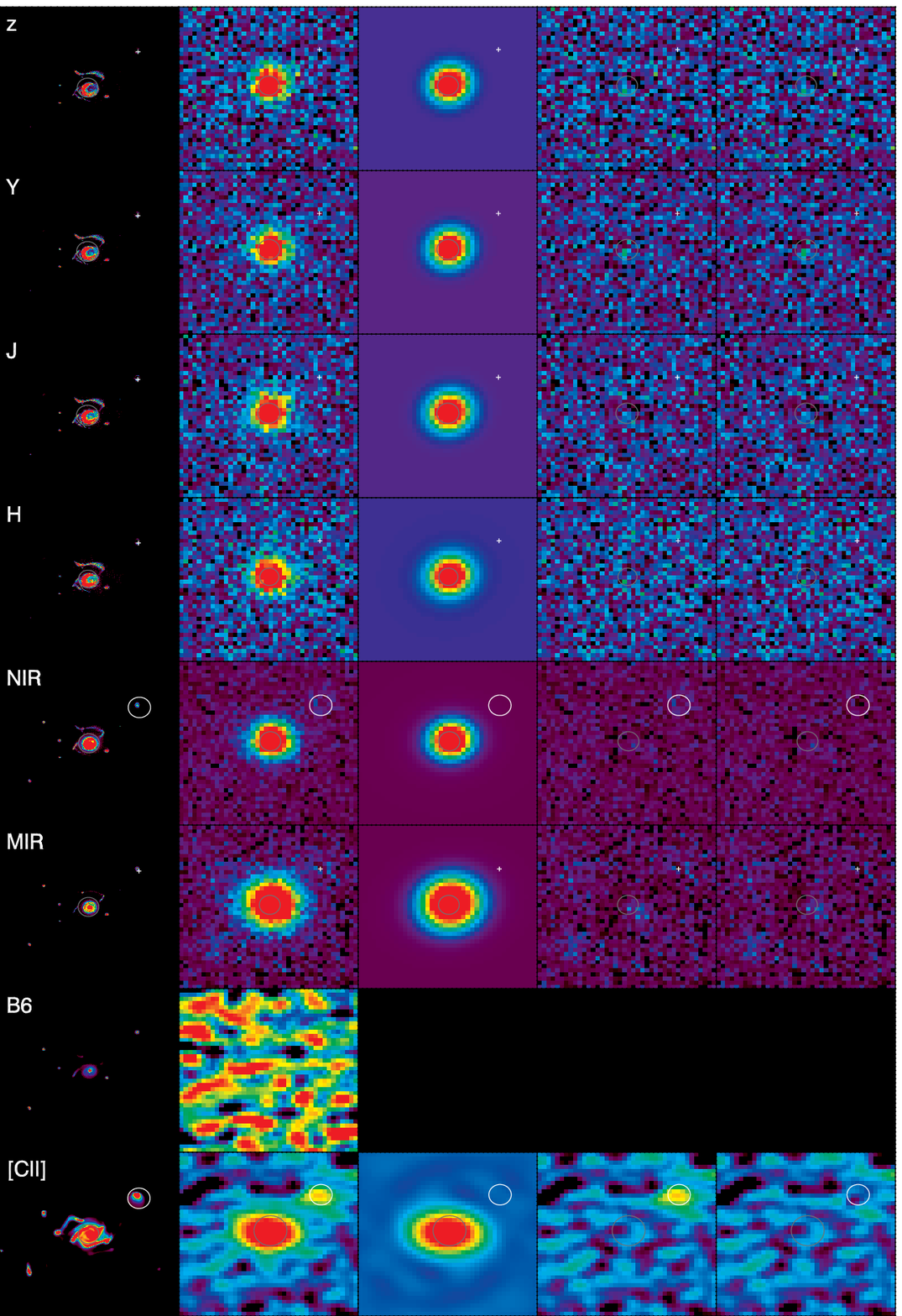}
\caption{Mock observations for the merger with resolution 0.15''. Images and symbols are the same as in Figure \ref{fig:maps16_0.05}}
\label{fig:maps23_0.15}
\end{figure*}

\clearpage

\section{Estimates of the effective radii}
\label{app:profiles}

\begin{figure*}
\centering
\includegraphics[width=0.8\textwidth]{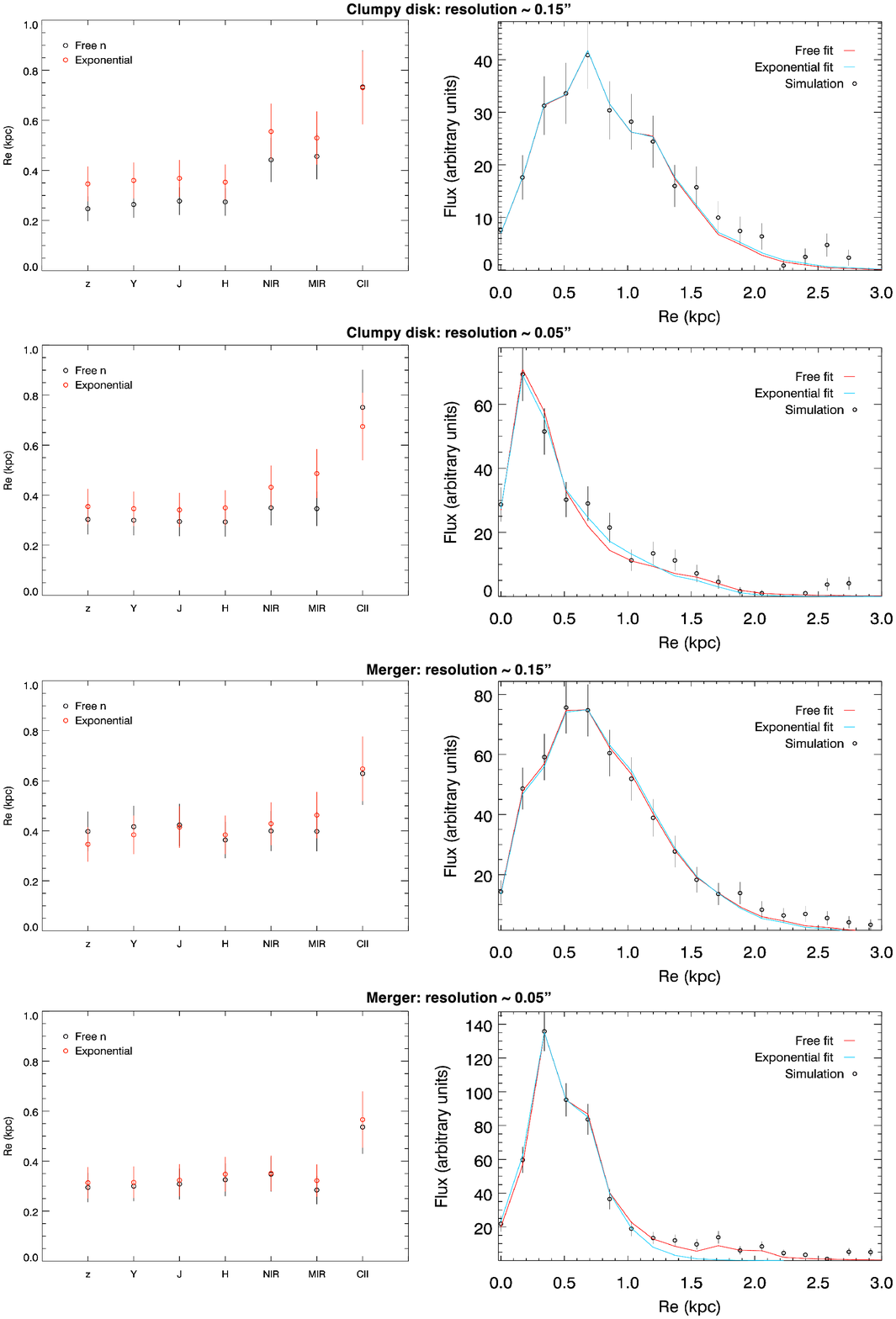}
\caption{Comparison of the effective radius estimated when fitting the simulations with a S\'ersic profile (with free S\'ersic index) and with an exponential profile. \textbf{Left column}: effective radius measured in different bands. Each plot (from top to bottom) refers to mock observations of the clumpy galaxy and the merger, with different angular resolution (0.15'' and 0.05''). The black (red) open circles indicate the effective radius obtained with a free S\'ersic (exponential) profile fit. \textbf{Right column}: we show the one-dimensional surface brightness profile of the galaxies (black open circles with Poissonian error bars), as observed in the MIR band. We also show the free S\'ersic (red curve) and exponential (cyan curve) profile fits. The effective radii measured with a free S\'ersic and an exponential profile are consistent within the uncertainties and there are no systematic trends with the observing band.}
\label{fig:profiles}
\end{figure*}

\bsp	
\label{lastpage}
\end{document}